\documentclass[final,authoryear,5p,times,twocolumn]{elsarticle}
\usepackage{graphicx,longtable}
\usepackage{natbib}
\usepackage{slashbox}
\usepackage{amssymb}
\usepackage{amsthm}
\usepackage{color}
\journal{New Astronomy}
\usepackage{lipsum}

\newcommand{\cf}{{\textrm cf.\,}}
\newcommand{\eg}{{\textrm e.g.\,}}
\newcommand{\ie}{{\textrm i.e.\,}}
\newcommand{\msun}{\mbox{\rm $M_{\odot}$~}}

\newcommand{\arcs}{\hbox{$^{\prime\prime}$}}
\newcommand{\arcm}{\mbox{$^{\prime}$}}

\newcommand{\ra}{\mbox{$\alpha_{2000}$}}
\newcommand{\dec}{\mbox{$\delta_{2000}$}}

\newcommand{\degree}{\mbox{$^{\circ}$}}

\newcommand{\hii}{\mbox{H{\sc ii}~}}

\def\ang{\AA\,}                           

\def\farcs{\hbox{$.\!\!^{\prime\prime}$}}

\def\gsim{\;\rlap{\lower 2.5pt\hbox{$\sim$}}\raise 1.5pt\hbox{$>$}\;}
\def\lsim{\;\rlap{\lower 2.5pt\hbo time lag between the formation x{$\sim$}}\raise 1.5pt\hbox{$<$}\;}
\def\la{\mathrel{\hbox{\rlap{\hbox{\lower4pt\hbox{$\sim$}}}\hbox{$<$}}}}
\def\ga{\mathrel{\hbox{\rlap{\hbox{\lower4pt\hbox{$\sim$}}}\hbox{$>$}}}}

\def\arcmin{\hbox{$^\prime$}}

\def\fdg{\hbox{$.\!\!^\circ$}}

\def\farcs{\hbox{$.\!\!^{\prime\prime}$}}

\begin{document}
\begin{frontmatter}
\title{Deep optical survey of the stellar content of Sh2-311 region}
\author[ari]{Ram Kesh Yadav\corref{cor}}
\ead{ramkesh@aries.res.in}
\cortext[cor]{Corresponding author}
\author[ari]{A. K. Pandey}
\author[ari]{Saurabh Sharma}
\author[iia]{J. Jose}
\author[jap]{K. Ogura}
\author[kiso]{N. Kobayashi} 
\author[lam]{M. R. Samal}  
\author[ari, tai]{C. Eswaraiah} 
\author[kum]{H. C. Chandola}   
\address[ari]{Aryabhatta Research Institute of observational sciencES (ARIES), Manora Peak, Nainital 263 002, India}
\address[iia]{Indian Institute of Astrophysics, Koramangala, Bangalore 560 034, India}
\address[jap]{Kokugakuin University, Higashi, Shibuya-ku, Tokyo 150-8440, Japan}
\address[kiso]{Kiso Observatory, School of Science, University of Tokyo, Mitake, Kiso-machi, Kiso-gun, Nagano-ken 397-0101, Japan}
\address[lam]{Aix Marseille Universit\'e, CNRS, LAM (Laboratoire d'Astrophysique de Marseille) UMR 7326, 13388 Marseille, France}
\address[tai]{Institute of Astronomy, National Central University, 300 Jhongda Rd, Jhongli, Taoyuan Country 32054, Taiwan}
\address[kum]{Department of Physics, Kumaun University, Nainital 263 002, India}


\linespread{1}

\begin{abstract}
The stellar content in and around Sh2-311 region have been studied using the deep optical observations as well as near-infrared (NIR) data from 2MASS. The region contains three clusters, viz. NGC 2467, Haffner 18 and Haffner 19. We have made an attempt to distinguish the stellar content of these individual regions as well as to re-determine their fundamental parameters such as distance, reddening, age, onto the basis of a new and more extended optical and infrared photometric data set. NGC 2467 and Haffner 19 are found to be located in the Perseus arm at the distances of 5.0 $\pm$ 0.4 kpc and 5.7 $\pm$ 0.4 kpc, respectively, whereas Haffner 18 is located at the distance of 11.2 $\pm$ 1.0 kpc. The clusters NGC 2467 and Haffner 19 might have formed from the same molecular cloud, whereas the cluster Haffner 18 is located in the outer galactic arm, \ie the Norma-Cygnus arm. We identify 8 class II young stellar objects (YSOs) using the NIR $(J - H)/(H - K)$ two colour diagram. We have estimated the age and mass of the YSOs identified in the present work and those by {\color{red} Snider et al. (2009)} using the $V/(V - I)$ colour-magnitude diagram. The estimated ages and mass range of the majority of the  YSOs are $\lesssim$1 Myr and $\sim$0.4 - 3.5 \msun, respectively, indicating that these sources could be T-Tauri stars or their siblings. Spatial distribution of the YSOs shows that some of the YSOs are distributed around the \hii region Sh2-311, suggesting a triggered star formation at its periphery. 
\end{abstract}


\end{frontmatter}

\section{Introduction}

The \hii region Sh2-311 (\ra = 07$^h$52$^m$26$^s$, \dec = -26\degree26\arcm12\arcs; $l$ = 243\fdg2, $b$ = 0\fdg4), also known as NGC 2467, is a part of the Puppis OB association. The region is located in ``Puppis Window" in the third quadrant of the Milky Way. The interstellar extinction towards the direction of Puppis Window is relatively low, hence the spiral arm structure as well as star formation activities at large distances can be studied.  The region is also interesting because it contains a number of young open clusters and nebular entities. The most conspicuous structure is Sh2-311, which is excited by an O6V type star HD 64315 ({\color{red} Crampton, 1971; Keenan \& Pitts, 1980; Walborn, 1982}). The region also contains clusters  Haffner 18 (H18) and  Haffner 19 (H19). The cluster H18 is elongated (cf. Section 3.1) along the galactic plane with an O6/O7V type star  as the most massive member and several late O and B type stars ({\color{red} Fitzgerald \& Moffat, 1974; Munari et al., 1998; Munari \& Carraro, 1996}), whereas the cluster H19 has a  B1.5V star as the most massive star ({\color{red} Lod\'{e}n, 1965; Moreno-Corral et al., 2002}).  The cluster H19 also contains two B2V, one B3V and two late B type stars. The entire region has many early-type stars, \eg\rm, HD 64568 (O3V), CD $-$26 5129 (O7), and CD $-$26 5126 (B0.5V), whose association  with the region is not known (see \eg\rm, {\color{red} Feinstein \& V\'{a}zquez, 1989; Munari \& Carraro, 1996; Munari et al., 1998}).

Several studies based on photometric and spectroscopic observations yielding conflicting results on distance estimation have already been carried out. The H18 region is a confusing one as it is reported to have two stellar groups, (viz. H18(1) and H18(2); \cf {\color{red} V\'{a}zquez et al., 2010}). The distance estimate to the H18 region varies from 6.3 to 8.4 kpc, whereas the distances to two stellar groups Haffner 18(1) and  Haffner 18(2) have been estimated as 4.5 $\pm$ 0.4 kpc and 9.5 to 11.4 kpc, respectively (\cf {\color{red} V\'{a}zquez et al., 2010}). The available distance estimation in the literature for H19 varies from 5.1 to 6.9 kpc (see {\color{red} V\'{a}zquez et al., 2010}). {\color{red} Feinstein \& V\'{a}zquez (1989)} reported that Sh2-311 (NGC 2467) is located at a distance of 4.1 kpc,  whereas {\color{red} Balser et al. (2011)} have estimated a kinematic distance of 5.3 kpc.  The conflicting distance estimates of the sub-regions  of the complex make it difficult to establish association among the sub-regions. Reliable estimation of distance is very important to study the star formation scenario in the region. 

We used deep optical observations of a wide field around Sh2-311 region and low resolution spectroscopy of a few bright stars to estimate the distances of the sub-regions as well as  to understand the stellar content in the region.  In addition to these observations we also used the available near-infrared (NIR) data from Two Micron All Sky Survey (2MASS; {\color{red} Cutri et al., 2003}), to identify young stellar objects (YSOs) associated with the region, which consequently used to understand the star formation scenario in the region.

\section {\label{wfi_obs} Observations and data reduction}

The data used in this work are a combination of archived optical $UBVI$ images taken with the Wide Field Imager (WFI) camera at the 2.2-m MPG Telescope of the European Southern Observatory (ESO) (for details see {\color{red} Gamen et al., 2006}) and NIR data from the 2MASS Point Source catalogue ({\color{red} Cutri et al., 2003}) available on the WEB. Low resolution spectroscopic observation of three stars were also taken using the 2.0-m Himalayan Chandra Telescope (HCT) of Indian Astronomical Observatory (IAO), Hanle.

\subsection{\label{opt_phot} Optical photometry}

\begin{figure*}[!htbp]
\centering
\includegraphics[width=10cm]{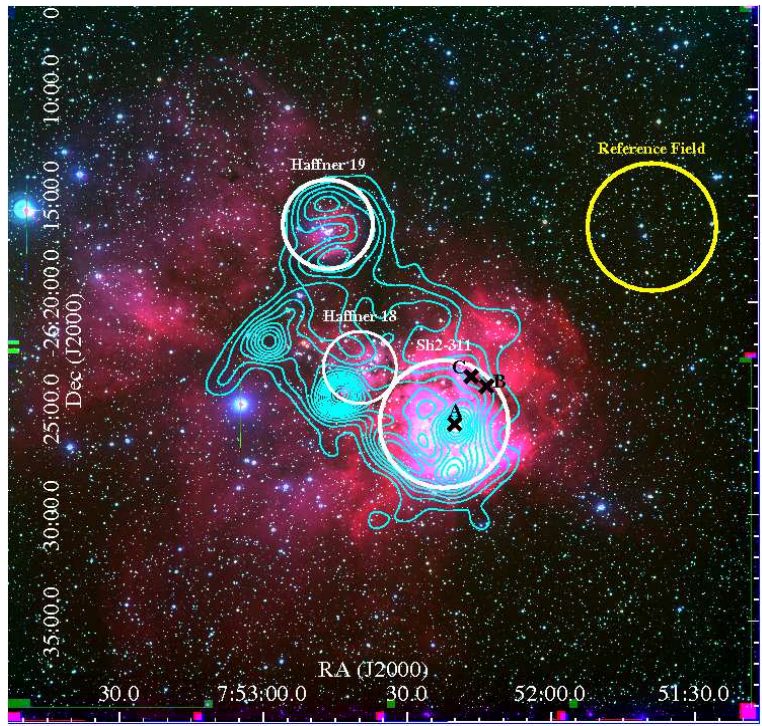}
\caption{The 34x33 arcmin$^2$ color composite image obtained by using $B$ (blue), $V$ (green) and $H_{\alpha}$ (red) filter images. The circles represent the estimated boundary of the clusters (cf. Section \ref{struct_clus}). The area marked by the yellow circle near the upper-right side of the image is the reference field for the optical data. The stars having spectroscopic observations are marked with crosses. MSX A-band contours are also overplotted. The isophot contour levels at 8.28 $\mu$m are 80, 75, 65, 60, 50, 40, 30, 25, 15 and 1 percent of the peak value.}
\label{op_cc_image} 
\end{figure*}

The archived optical observations of Sh2-311 region used in this study were taken using the WFI camera (PI: Rubio, M.) mounted at the Cassegrain focus of the ESO 2.2-m Telescope. This instrument consists of 4 $\times$ 2 mosaic CCDs of 2048 $\times$ 4096 square pixels with a plate scale of 0.238 arcsec/pixel. Each chip covers an area of $\sim$8.$\arcmin$12 $\times$ 16.$\arcmin$25, while the full field of view (FOV) of the camera is 34$\arcmin$ $\times$ 33$\arcmin$. The observations were retrieved from the ESO's Science Archive Facility.  Details of the observations are given in Table \ref{obslog}. The frames were taken under slightly variable seeing conditions with full width half maximum (FWHM) of the point-spread function varying from 0.81 to 1.69 arcsec. The observed region is shown in Fig. \ref{op_cc_image}.

The first stage of the data reduction was carried out using  a mosaic specific {\it MSCRED} package in IRAF. The instrumental electronic bias of the images was subtracted using the overscan region and then the images were trimmed to remove the overscan region. The contribution of zero level exposure patterns was estimated by combining a number of shutter closed, zero exposure frames and the same was subtracted from the flat field and object frames. The {\it ZEROCOMBINE} task of {\it MSCRED} was used to combine the zero exposure frames.   Flat fielding for each filter was performed using a set of sky flat fields scaled to the mode value of all chips combined into a master flat using the {\it MSCRED}'s {\it FLATCOMBINE} task.  Bad pixel templates were created by running the {\it CCDPROC} task on the science frames and these defects were removed by interpolation from nearby good pixels. The geometric shifts  were applied to align the dithered exposures using the standard procedures in the IRAF (\cf the NOAO mosaic data reductions link\footnote{http://www.noao.edu/noao/noaodeep/ReductionOpt/frames.html}).

The photometry was performed using IRAF's DAOPHOT-II package ({\color{red} Stetson, 1987}). Since  the region is crowded  as well as has a large amount of diffuse emission in optical bands, the point-spread function (PSF) photometry was performed. The PSF was constructed using 50 well isolated stars free from the nebulosity and distributed  throughout the frames.  The task {\it PSTSELECT} and {\it PSF} were used to select the  stars and construct the  PSF, respectively.  The source detection on the final images was carried out  using  {\it DAOPHOT-II/ALLSTAR} photometric routines described in {\color{red} Stetson (1987)}. The observations were standardized using the recent photometry of H18 and H19 regions by {\color{red} V\'{a}zquez et al. (2010)}.

\begin{figure*}[!htbp]
\centering
\includegraphics[width=10cm, angle=270]{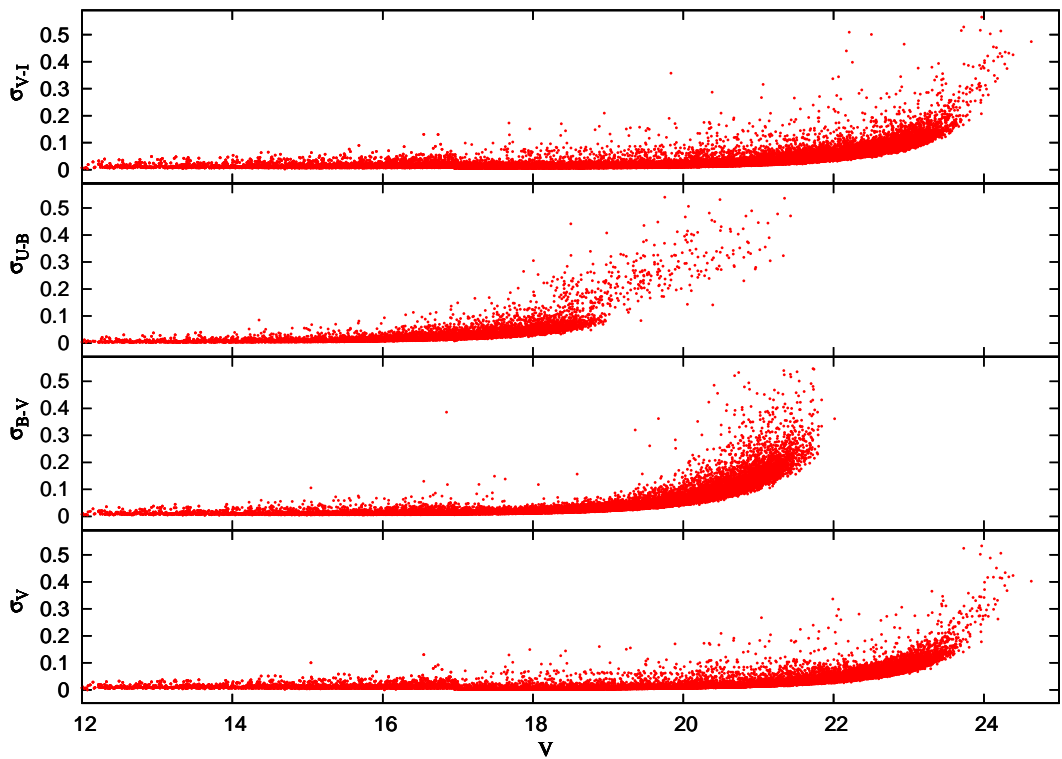}
\caption{\label{sigp} Typical DAOPHOT error in the magnitude and colours as a function of V magnitude.}
\end{figure*}

Calibration of the instrumental magnitude to the standard system was done using the following photometric calibration equations. \\
\[ (U - B) = m_1*(u - b) + c_1  \]
\[ (B - V) = m_2*(b - v) + c_2  \]
\[ (V - I) = m_3*(v - i) + c_3  \]
\[  V = v + m_4*(V - I) + c_4   \]

where $u, b, v, i$ are the instrumental magnitudes corrected for atmospheric extinction and $U, B, V, I$ are the standard magnitudes. $m_1, m_2, m_3$ and $m_4$ and $c_1, c_2, c_3$ and $c_4$ are colour coefficients and zero-point coefficients respectively. We used the mean values of the extinction taken from the WFI homepage\footnote{http://www.eso.org/sci/facilities/lasilla/instruments/wfi/inst/zeropoints/}. The values of the colour coefficients and the zero point coefficients in various bands are given in Table \ref{col_coef}. When brighter stars were saturated in long exposure $V$, and $I$ frames, their magnitudes were taken from short exposure frames. In the case of $U$ and $B$ pass-bands we have used only short exposure frames. The $U$ and $B$ band observations of brighter stars were used only to estimate the reddening in the region, whereas in the case of $V$ and $I$ pass-bands we used deep exposure in addition to short exposure observations to study the faint stellar content of the region. The typical DAOPHOT errors in the magnitude and colour as a function of $V$ magnitude are shown in Fig. \ref{sigp}. It is apparent that we can go up to $V$ $\simeq$ 22 mag with corresponding error $<$ 0.1 mag. This is about 2 mag deeper than the most recent CCD observation of the region (cf. {\color{red} V\'{a}zquez et al., 2010}). Stars brighter than 12 magnitude were saturated even in short exposure frames and were taken from the literature wherever available. 

Astrometric calibration that relates the image pixels to equatorial celestial coordinates was done by using  {\it MSCRED} package. For a mosaic CCD each extension requires its own world coordinate system (WCS) to account for the relative orientations of the CCDs as well as for the  optical distortions from the focal plane to the sky. {\it MSCRED}'s tasks {\it MSCSETWCS}, {\it MSCCMATCH}, {\it MSCGETCAT}, {\it MSCTVMARK}, {\it MSCZERO} were used to update the  WCS information of the WFI images. We used {\it WCSTOOLS}'s {\it XY2SKY} task to convert pixel coordinates to equatorial celestial coordinates. We  compared our astrometry with that of the stars in the USNO A2 catalogue and the mean and standard deviation in RA and Dec are $-$0.06$\pm$0.48 arcsec and $-$0.09$\pm$0.37 arcsec, respectively.

\begin{figure*}[!htbp]
\includegraphics[trim=0.00cm 0.0cm 0cm 0.0cm, clip, width=15cm]{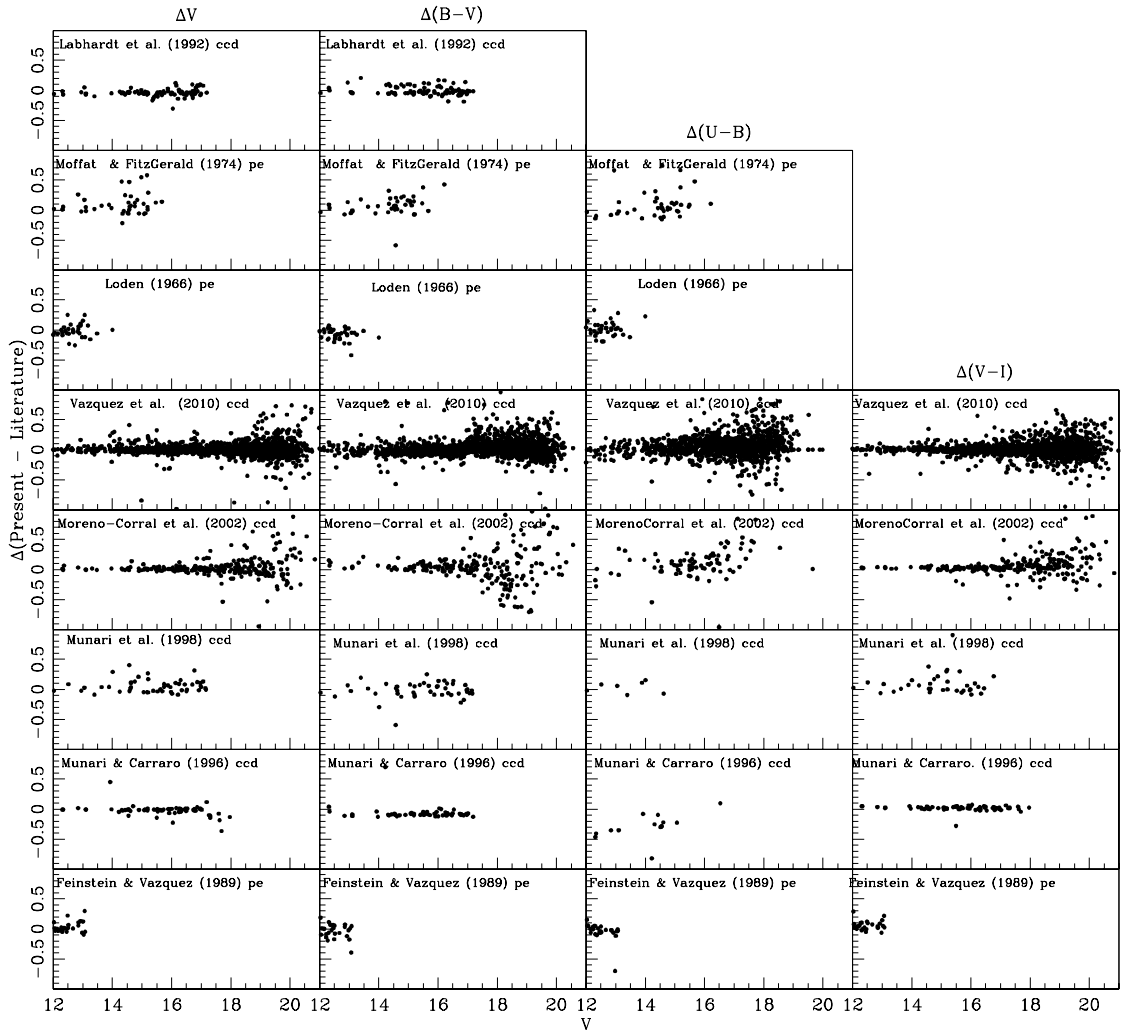} 
\caption{\label{phot_comp} Comparison of present photometry with the photometry available in the literature. The $\Delta$ represents the difference between the present and literature values. (This figure is available in electronic version only). The ccd and pe represent CCD and photoelectric observations, respectively.}
\end{figure*}

 While selecting a reference field at the corner of the image we noticed a systematic shift in the ($U - B$) colour, hence we did not consider the data having \dec $<$ $-$26\degree10\arcm56\arcs and \dec $>$ $-$26\degree33\arcm38\arcs in the present study. We selected reference field centered at \ra = 07$^h$51$^m$39.$\!^s$36, \dec = $-$26\degree16\arcm29\farcs73 is shown in Fig. \ref{op_cc_image} and has a radius of 3\arcmin. A sample of optical photometric data containing RA, Dec, magnitude, and colours of the stars in Sh2-311 region is given in Table \ref{op_data}. Complete data is available in electronic form only and can be obtained from the authors. A comparison of the present photometry with those  available in the literature is shown in Fig. \ref{phot_comp} and the results are given in Table \ref{photdata}. The difference $\Delta$(present $-$ literature) are computed after applying a 3$\sigma$ clipping to the photometric residuals. The comparison suggests that the present photometry is in fair agreement with the previous available photometries.

Study of luminosity function/mass functions (LFs/MFs) requires necessary corrections in the data sample to take into account the incompleteness that may occur due to various reasons (\eg, crowding of the stars). We used the {\it ADDSTAR} routine of DAOPHOT II to determine the completeness factor (CF). The procedures have been outlined in detail in our earlier works (see \eg, {\color{red} Pandey et al., 2008}). Briefly, we  randomly added artificial stars to  $V$ and $I$ images in such a way that they have similar geometrical locations but differ in $I$ brightness according to mean $(V - I)$ colour (1.45 mag) of the data sample. The luminosity distribution of artificial stars was chosen in such a  way that more number of stars were inserted towards the fainter magnitude bins. The frames were then reduced using the same procedure  as used for the original frames. The ratio of the number of stars recovered to those added in each magnitude interval gives the CF as a function of magnitude. The minimum value of the CF of the pair (\ie $V$ and $I$ bands) for the cluster regions and the nearby reference field region (\ra = 07$^h$51$^m$39.$\!^s$36, \dec = $-$26\degree16\arcm29\farcs73) are given in  Table \ref{cft}. As expected the incompleteness increases with the magnitude, however the data are $\sim90\%$ complete up to $V \sim$21 magnitude.

\subsection{Optical spectroscopy}

Low-resolution optical spectroscopy for three bright sources in Sh2-311 region were carried out by using  the Himalayan Faint Object Spectrograph and Camera (HFOSC) of the 2.0-m HCT. The log of spectroscopic observations is given in Table \ref{log}. These three stars  (\#A, \#B and \#C; \cf Table \ref{spec_mem}) are the optically bright sources lying within $\sim$3$\arcmin$ radius of   Sh2-311. The spectra in the wavelength range 3800$-$6840 \ang with a dispersion of 1.45 \ang/pixel were obtained by using the low-resolution Grism 7 with a slit of 2 arcsec width. The spectra were bias subtracted, flat fielded and then one-dimensional spectra were extracted by using the optimal extraction method in IRAF. Wavelength calibration of the spectra was done by using an Fe$-$Ar lamp source. Spectrophotometric standard star Feige 34 ({\color{red} Oke, 1990}) was observed on the same night for flux calibration.

\subsection{NIR photometric data}

NIR  $JHK_s$ data for point sources for the region have been obtained from the 2MASS Point Source Catalogue ({\color{red} Cutri et al., 2003}). To improve the photometric accuracy, we used photometric quality flag (ph${\_\rm qual}$ = AAA), which gives an S/N $\ge$ 10 and a photometric uncertainty $<$ 0.10 mag. This selection criterion ensures best quality detection in terms of photometry and astrometry (\cf {\color{red} Lee et al., 2005}). 

\section{\label{struct_clus} STRUCTURE OF THE CLUSTERS}

\subsection{Two-dimensional surface density distribution}

The initial stellar distribution of star clusters may be governed by the structure of the parental molecular cloud as well as how star formation proceeded in the cloud (see \eg {\color{red} Chen et al., 2004; Sharma et al., 2006}).  Later evolution of the cluster may then be governed by internal gravitational interaction among member stars and external tidal forces due to the Galactic disc or giant molecular clouds.

\begin{figure*}[!htbp]
\centering
\includegraphics[trim=0cm 0cm 0.2cm 0cm, clip, width=10cm]{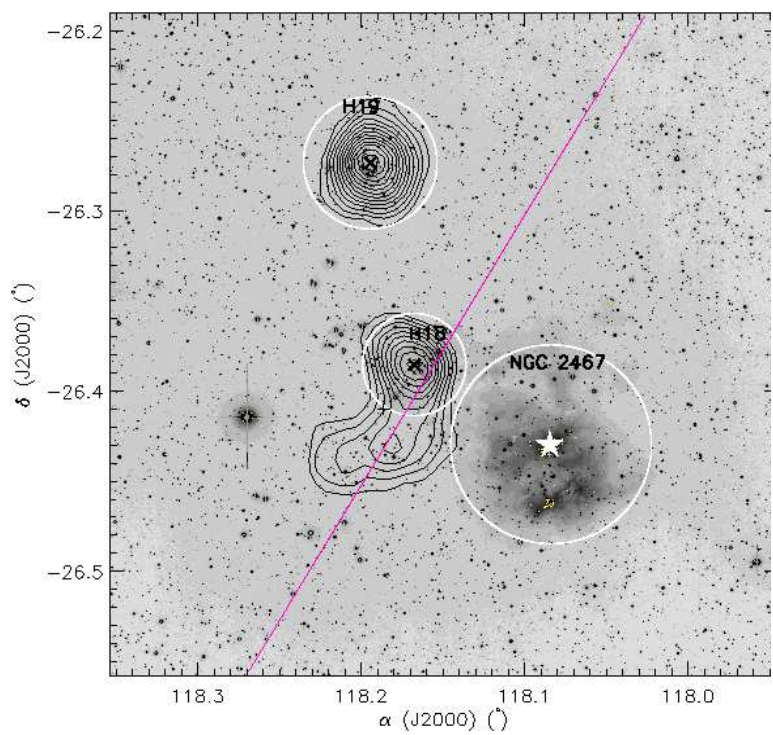}
\caption{\label{op_isd_contours} Isodensity contours (thick black curves) of the WFI $I$-band data shown with a grid size of 45$\times$45 arcsec$^2$. The lowest contour is plotted at 1$\sigma$ level above background. The cross marks represent the cluster centers. The white star symbol represents the location of the ionizing source HD 64315. The estimated radial extent of the clusters H19, H18 and Sh2-311 are also shown by white circles. The  galactic equator is shown by a line.}
\end{figure*}

To study the morphology of the clusters in the region, we generated isodensity  contours for stars detected  in $I$ band ($\sigma$ $<$ 0.1 mag)  and  the same are shown in Fig. \ref{op_isd_contours}. The contours are plotted above 1$\sigma$  of the background level as estimated from the reference field. The surface density distribution of the optical data reveals two prominent groups of stars at \ra = 07$^h$52$^m$39$.\!^s$9, \dec = $-$26\degree23\arcm07\farcs26 and \ra = 07$^h$52$^m$46.$\!^s$7, \dec = $-$26\degree16\arcm23\farcs73, corresponding to H18 and H19, respectively. The stellar surface density contours for H18 reveals an elongated structure for the cluster. The cross symbols in Fig. \ref{op_isd_contours} represent the adopted center of the respective clusters.  The Sh2-311 region does not show any stellar density enhancement, however we can clearly estimate the extent of the ionized gas from the extent of the nebulosity.

\subsection{Radial stellar surface density and cluster size}

\begin{figure}[!htpb]
\centering
\hbox{
\includegraphics[angle=270,width=8cm]{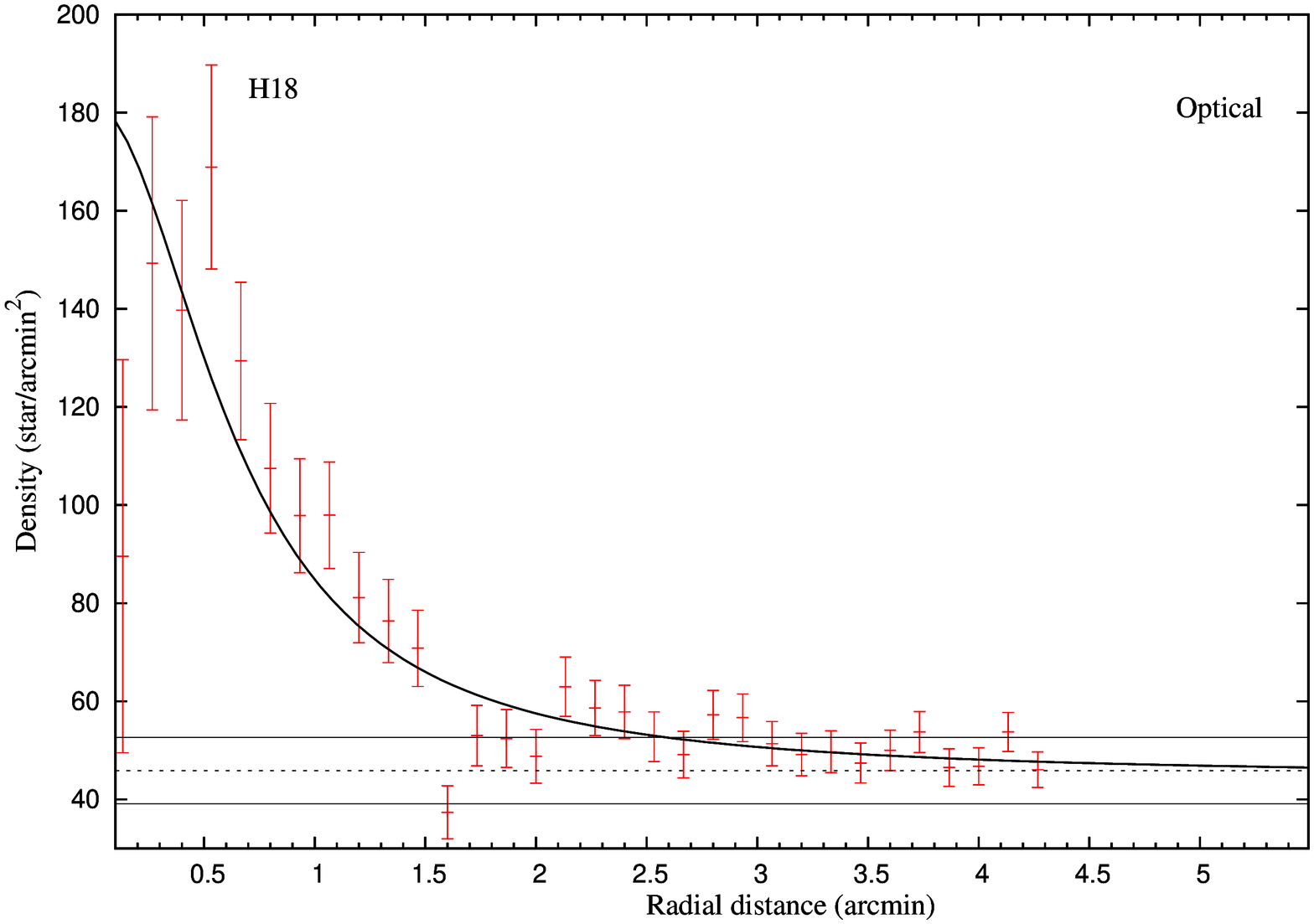} 
\includegraphics[angle=270,width=8cm]{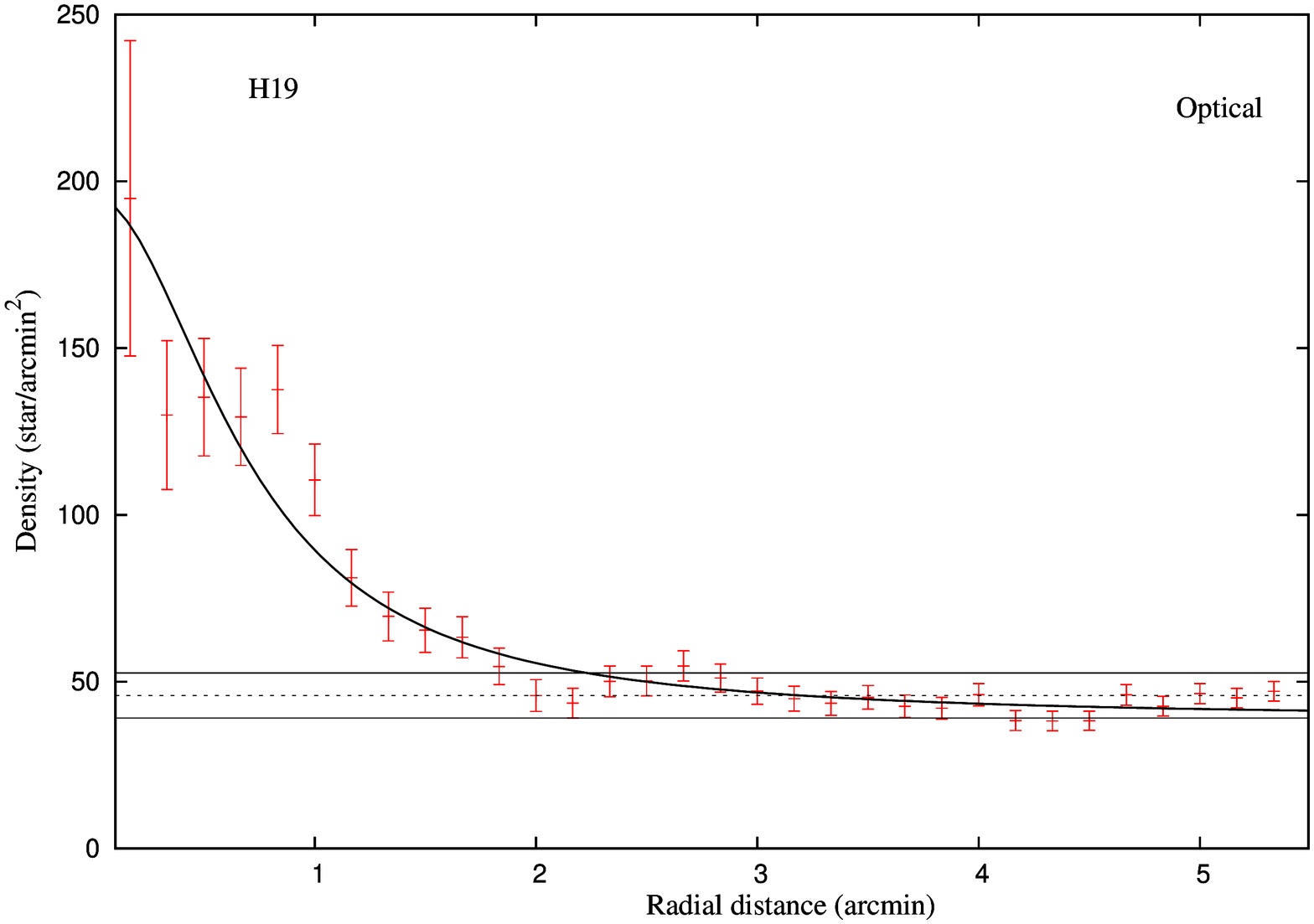}
}
\caption{\label{rdp} Stellar density profiles of H18 and H19 as a function of radial distance from the adopted cluster center. The continuous curve shows the least square fit of King's  {\color{red} (King, 1962)} profile to the observed data. The dotted line represents the mean density level of field stars as estimated from the reference field. The continuous lines represent the 1$\sigma$ error in this mean density level.}
\end{figure}

The radial extent of the cluster is one of important parameters used to study its dynamical state. To investigate this we have derived the radial density profile (RDP) for H18 and H19 cluster regions using the optical ($V \le$ 20 mag) photometric data. Sources were counted in concentric annulus of 10-arcsec width around the cluster center and the counts were then normalized by their respective area. These density (normalised counts) plotted as a function of radius is called RDP and is shown in Fig. \ref{rdp}. The error bars are derived by assuming that the number of stars in each annulus follows the Poisson statistics.

The radial extent of the clusters $(r_{cl} )$ is defined as the point where the cluster RDP merges with the field star density. The observed RDPs for H18 and H19 seem to merge with the field star density within 1$\sigma$ error at radial distance of $\sim$1.8 arcmin and $\sim$2.0 arcmin, respectively. {\color{red} Haffner (1957)} estimated the apparent diameter of H19 as 1.8 arcmin. However, in a recent study {\color{red} V\'{a}zquez et al. (2010)} estimated the extents of H18 and H19 as $\sim$3.6 and $\sim$3.0 arcmin, respectively, which are significantly higher than the values estimated in the present work. To parametrize the RDP, we fitted the observed RDP with the empirical model of {\color{red} King (1962)}, which is given by
\[
\rho(r) = \frac{f_0}{1+\left(\frac{r}{r_c}\right)^2}
\]
where $r_c$ is the core radius at which surface density $\rho(r)$ becomes half of the central density, $f_0$. The best fit to the observed RDPs obtained by a $\chi^2$ minimization technique is also shown in Fig. \ref{rdp}. The core radii thus estimated from the RDPs for clusters H18 and H19 are 0.64 $\pm$ 0.14 and 0.69 $\pm$ 0.11 arcmin, respectively. It is worthwhile to mention that King's profile for H18 and H19 seems to merge with the mean density of the field at $\sim$4 arcmin and $\sim$3 arcmin, respectively, which indicate a larger extents for the clusters as suggested by {\color{red} V\'{a}zquez et al. (2010)}. However, to avoid contamination due to field population we preferred the radial extents for H18 and H19 as 1.8 arcmin and 2.2  arcmin respectively. Since the Sh2-311 region is a heavily obscured region, the optical observations of the region are not sufficient to draw the RDP. We visually identified the extent of the ionized region using the $V$ band image as $\sim$3 arcmin.

\section{Analysis of the data}

\subsection{\label{redd} Reddening and extinction law}

\begin{figure*}[!htbp]
\centering
\includegraphics[trim=1.3cm 4.9cm 0.0cm 0cm, clip, width=15cm]{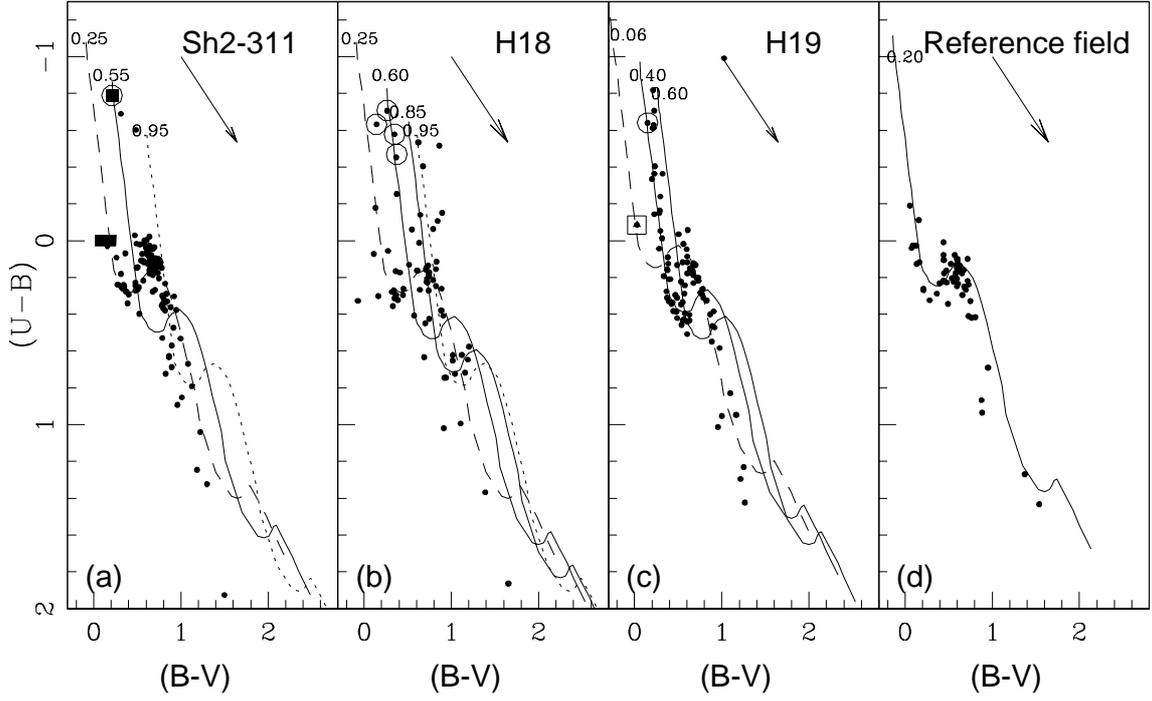} 
\caption{\label{op_tcd} $(U-B)/(B-V)$ TCD for stars in the Sh2-311, H18, H19 and nearby reference field regions. The filled squares represent stars having spectroscopic observations. The curves represent the ZAMS by {\color{red} Schmidt-Kaler (1982)} shifted along the reddening slope of 0.72 (shown by an arrow). The value of the $E(B - V)$  is marked at the top of the shifted ZAMS.  The dashed curves in the panels (a) and (b)represent the ZAMS which is reddened by $E(B-V)$ = 0.25 mag to match with the probable foreground population. The short-dashed curves represents the probable background Norma-Cygnus population. Stars marked with open circles are O/B type stars in the region. The star marked with open square in H19 region is a foreground star (cf. Section \ref{redd})}.
\end{figure*}

The extinction towards a star forming region (SFR) can arise due to the following two causes: (1) the general interstellar medium in the foreground [$E(B - V)_{min}$], and (2) the localized cloud associated with the SFR [$E(B - V)_{local} = E(B - V)_{max} - E(B - V)_{min}$], where $E(B - V)_{max}$ is the observed reddening of the star embedded in the local cloud. The Sh2-311 region,  as clearly evident from Fig. \ref{op_cc_image} and \ref{op_isd_contours} is still embedded in the parent molecular cloud and can have a variable extinction. {\color{red} V\'{a}zquez et al. (2010)} have pointed out that the clusters H18 and H19 are located in a complex region. To estimate the extinction  towards the Sh2-311, H18 and H19 regions, we used  $(U - B)/(B - V)$ two-colour diagram (TCD), which are shown in Fig. \ref{op_tcd}. In Fig. \ref{op_tcd} we have also plotted TCD for the reference field (cf. Section 2.1). Since the region Sh2-311 is very young, the TCDs show a scatter distribution of stars. The TCDs  also indicate a significant amount of contamination due to field stars.  The reddening $E(B-V)$ towards the reference field region is found to be $\sim$0.20 mag. The probable foreground stars towards the Sh2-311 and H18  regions follow  the intrinsic zero-age main-sequence (ZAMS) by {\color{red} Schmidt-Kaler (1982)} reddened by $E(B - V)$ = 0.25 mag (dashed line) along a normal  reddening vector [\ie $E(U - B)/E(B - V)$ = 0.72]. In Fig. \ref{op_tcd}, the continuous line represents the intrinsic ZAMS  shifted along the normal  reddening vector to match the observed array of probable cluster members. The Sh2-311 region reveals a minimum reddening in the region as  $E(B-V)_{min}$ $\simeq$ 0.55 mag. The brightest star (HD 64315) of Sh2-311 region is O6V star (\cf Section 4.2). Its intrinsic $(B - V)_0$ colour (-0.32 mag) and observed colour $(B - V)$= 0.21 mag also yields $E(B - V)$=0.53 mag.  H18 and H19 regions seems to show a variable reddening with $E(B - V)_{min}$ and $E(B - V)_{max}$ of 0.60 mag to 0.85 mag and 0.40 mag to 0.60 mag, respectively. In the case of H19, {\color{red} Munari \& Carraro (1996)} measured a variable reddening of 0.5 mag with a mean value of $E(B - V)$=0.44, whereas {\color{red} V\'{a}zquez et al. (2010)} estimated $E(B - V)$=0.38$\pm$0.02. {\color{red} Munari et al. (1998)} on the basis of spectroscopic and photometric observations of H18 region derived a mean colour excess  $E(B - V)$=0.62, whereas {\color{red} V\'{a}zquez et al. (2010)} for H18 found a differential reddening in the region ranging from  $E(B - V)$ $\sim$0.6 to 1.0 mag. The estimated values of reddening in the present work are in  fair agreement with the estimates available in the literature. A careful inspection of the TCDs indicates the presence of further reddened population towards the Sh2-311 and H18 region. The ZAMS shown by short-dashed line is further shifted to match the reddened sequence. This population may belong to the Norma-Cygnus arm  ({\color{red} Carraro et al., 2005; Pandey et al., 2006}). The $E(B - V)$ value for the background population comes out to be $\sim$0.95 mag. The dashed curve in the left and middle panel of Fig. \ref{op_tcd} represent the foreground population having reddening $E(B - V)$=0.25 mag. 

\begin{figure*}[!htbp]
\centering
\includegraphics[angle=270,width=15cm]{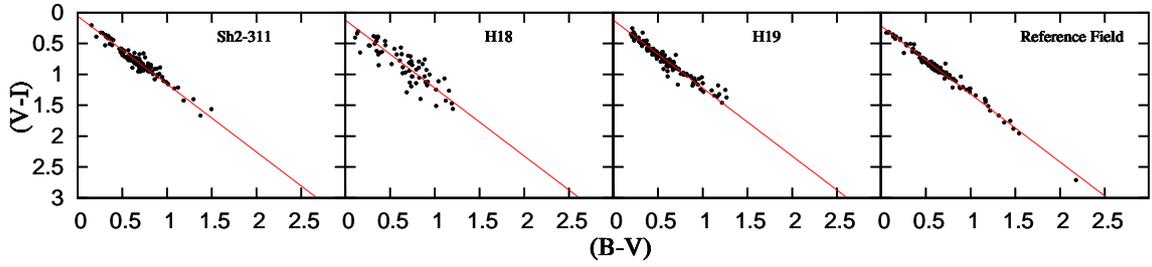}  
\caption{\label{op_tcdr} $(V - I)/(B - V)$ TCD for Sh2-311, H18, H19, and nearby reference field. Solid line represents normal reddening law.}
\end{figure*}

 The interstellar extinction and the ratio of total-to-selective extinction $R_V = A_V/E(B - V)$ in the cluster region are important parameters, which can influence the distance estimation of the cluster. To study the nature of the extinction law towards the cluster regions, we used $(V - I)/(B - V)$ TCDs as described by {\color{red} Pandey et al. (2003)} and references therein. The $(V - I)/(B - V)$ TCD provide an effective method for separating the influence of the normal extinction produced by the diffuse interstellar medium from that of the extinction due to the intra-cluster medium (\cf {\color{red} Chini \& Wargau, 1990; Pandey et al., 2003}). The TCDs for the Sh2-311, H18, H19 and reference field region are shown in Fig. \ref{op_tcdr}. The TCD of the reference field shows a well defined distribution manifesting a normal reddening law. The distribution of colour in the TCDs of the cluster regions show relatively increased scatter. The straight line in the TCDs show a normal reddening law indicating that the distribution of stars in  $(V - I)/(B - V)$ TCD follow the normal reddening. In the case of H18 {\color{red} V\'{a}zquez et al. (2010)} have concluded that although there was evidence that a small fraction of stars may be affected by anomalous extinction, the reddening towards the clusters region in general could be characterized by the normal value, i.e. $R_V$=3.1. In the case of H19 {\color{red} Moreno-corral et al. (2002)} have concluded that the photometry in the H19 region shows that a normal extinction law is suitable for the region. Hence we adopted a normal reddening law towards the Sh2-311, H18 and H19 region.

\subsection{\label{spec_cla} Spectral classification of the bright sources in the Sh2-311 region}

\begin{figure*}[!htbp]
\centering
\includegraphics[width=10cm]{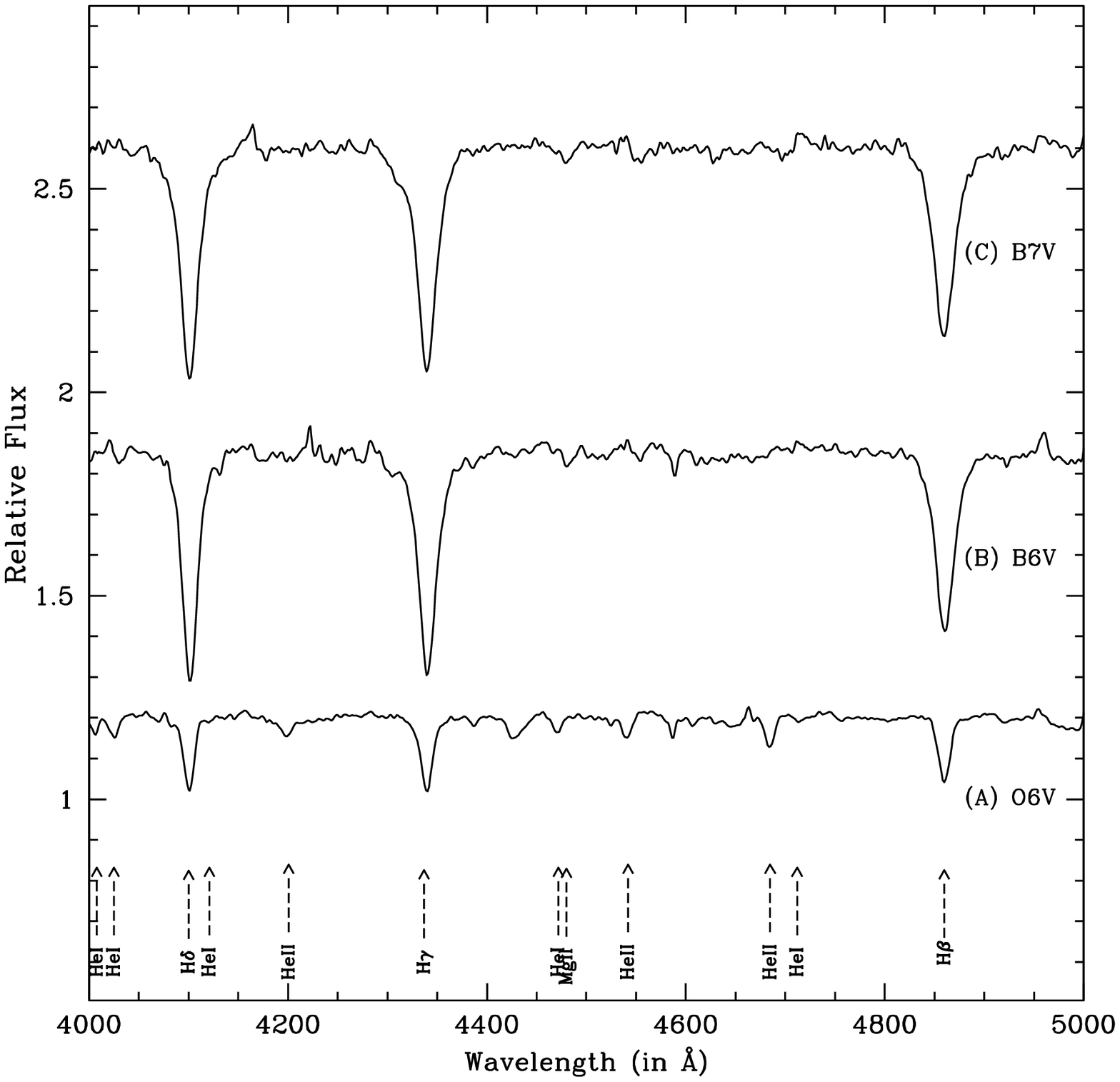} 
\caption{\label{spectra} Flux calibrated, normalized spectrum for the ionizing source as well as for the B6V and B7V type stars in the Sh2-311 region. The lines identified for the spectral classification are marked.}
\end{figure*}

We carried out low resolution spectroscopy of three optically bright sources in the Sh2-311 region. These objects  are marked in Fig. \ref{op_cc_image} as well as in  the $(U -B) / (B - V)$ TCDs and color-magnitude diagram (CMD) (see Figs. \ref{op_tcd}a and \ref{op_cmd}a). These three stars were selected based on their brightness, their location within 3 arcmin radius of Sh2-311 as well as their location close to the ZAMS on the CMD.  The flux calibrated, normalized spectra of these sources are shown in Fig. \ref{spectra} and the important stellar absorption lines are labeled.

The presence of neutral and/or ionized helium lines in absorption is indicative of O or early B type dwarfs, whereas these lines may be in emission in super giants. Star \#A shows HeI and HeII absorption lines. In the case of O type stars, the ratio of HeI 4471/HeII 4542 is a primary indicator of its  spectral type. This ratio is found to vary from $<$ 1 to $>$ 1 as we move from O5 to O7 and later types. The line strength of HeII 4686 gets weaker for  late O type stars and HeII 4686 is last seen in B0.5 type stars ({\color{red} Walborn \& Fitzpatrick, 1990}).  Star \#A has HeI 4471 \ang $<$ HeII 4542 \ang  and HeII 4686 \ang in absorption. The lack of NIII 4634-42 \ang  and CIII 4650 \ang in emission indicates  that this star is a main-sequence (MS) star. We classified star \#A using the ratio of the equivalent width (EW) of  HeI 4471 \ang  and HeII 4542 \ang. The ratio (\ie Log EW=Log[EW (HeI 4471 \ang)/EW (HeII 4542 \ang)]) is found to be $\sim$-0.16, implying that this star is likely to be of spectral type earlier than O7. Following the criteria given by {\color{red} Conti \& Alschuler (1971)} for spectral classification, we assign a spectral class of O6 $\pm$ 1  to star \#A. This star, being the brightest source in the region and also known as HD 64315, is the probable ionizing source of Sh2-311. The spectral type of this star has already been discussed in the literature as O6V ({\color{red} Crampton, 1971; Keenan \& Pitts, 1980; Walborn \& Fitzpatrick, 1990}), which is in agreement with our estimation.

In Fig. \ref{spectra} we have shown the flux calibrated, normalized spectra of stars \#B and \#C, which appear almost similar.  The much stronger Balmer lines in comparison to star \#A indicate that the stars \#B and \#C may be late B type stars. To classify B type stars we used the criteria by {\color{red} Jacoby et al. (1984)} and {\color{red} Torres-Dodgen \& Weaver (1993)}. The presence of weak HeI lines in absorption constraints the spectral types to be earlier than B5-B7.  Finally, the spectral type was assigned to these sources by visual comparison to the standard library spectra given in the literature ({\color{red} Jacoby et al., 1984}) and we assigned spectral classes  of B6 and B7 for stars \#B and \#C, respectively. However, because of low resolution observations, an uncertainty of $\pm$ 0.5 to 1 in the sub-class identification is expected.

We calculated the distances of these sources based on their assigned spectral classes. The absolute magnitudes and colors are taken from {\color{red} Koornneef (1983)}. The distances are calculated using  their observed magnitudes and colors both in optical and NIR (2MASS). The average distances thus calculated are  5.0 kpc, 1.7 kpc  and 1.9  kpc, respectively for star \#A, \#B and \#C.  The distance estimation for star \#A is in agreement with that obtained by using the optical CMD of Sh2-311 (see Section \ref{op_cmd1}). However, the results for stars \#B and \#C suggest that these should be foreground stars.

\subsection{\label{op_cmd1} Optical colour-magnitude diagrams: distance and age}

\begin{figure*}[!htbp]
\centering
\vbox{
\includegraphics[trim=0cm 4cm 0cm 0cm, clip, height=7cm]{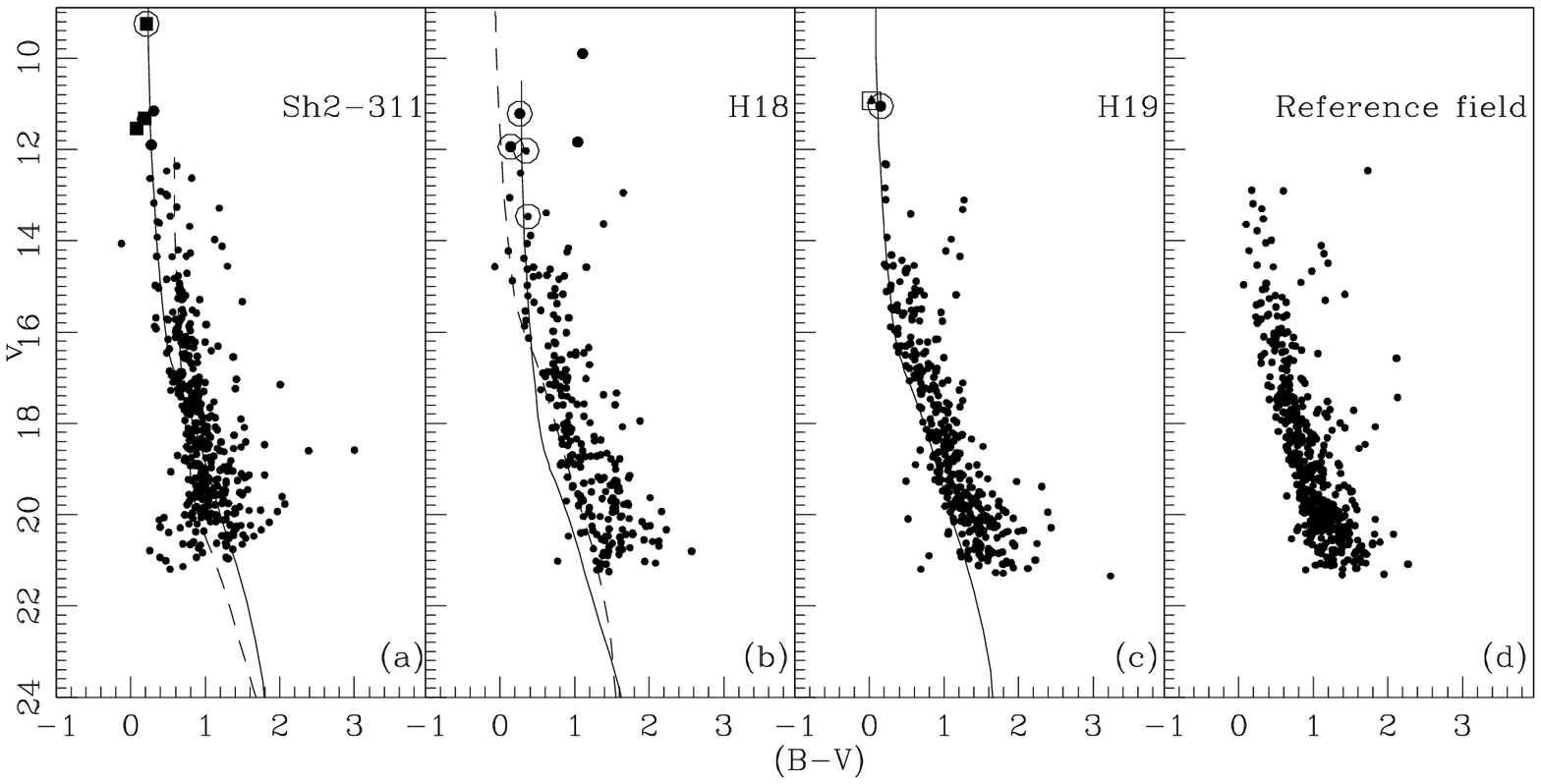} 
\includegraphics[trim=0cm 4cm 0cm 0cm, clip, height=7cm]{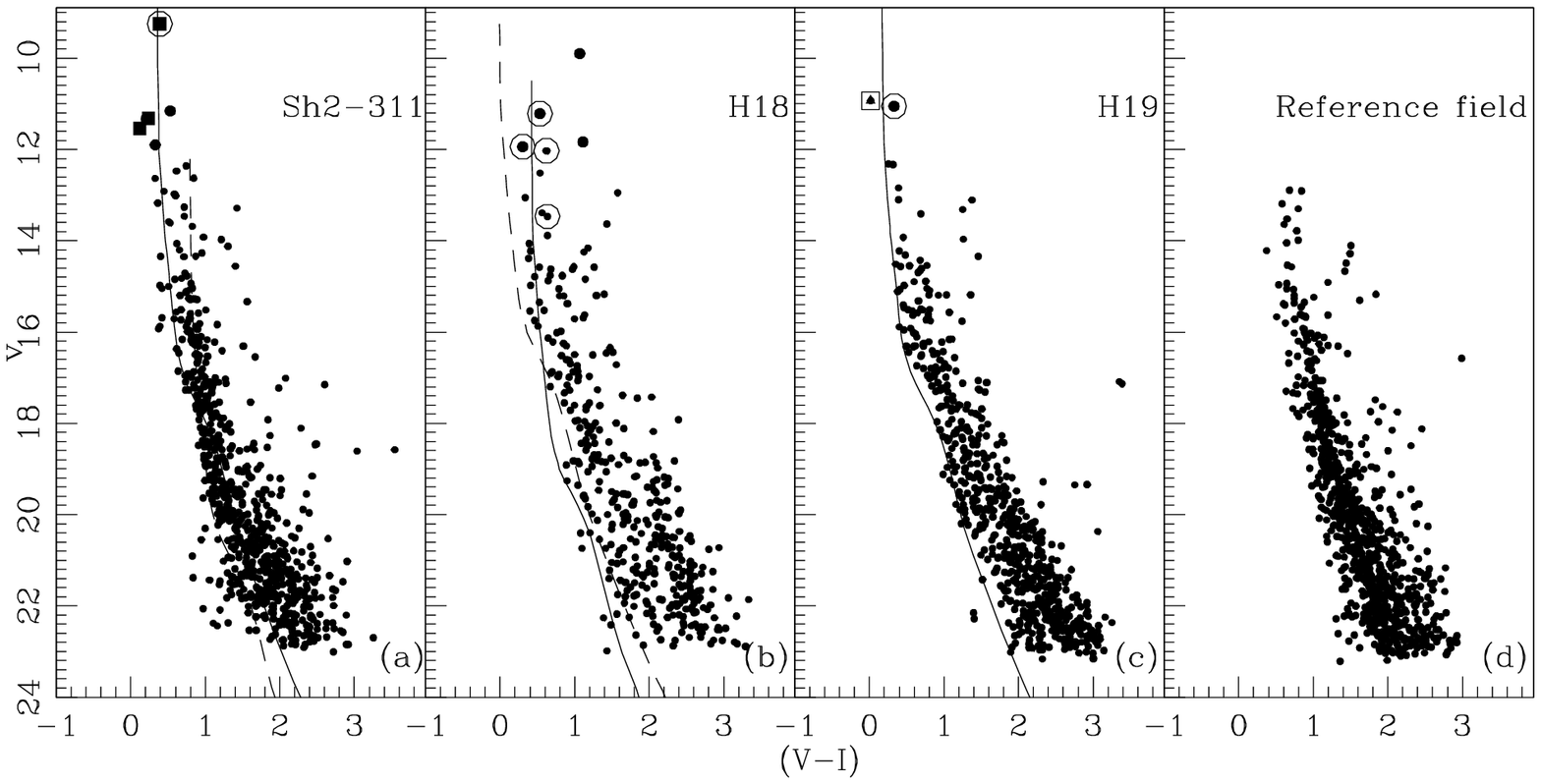} 
}
\caption{\label{op_cmd} $V/(B-V)$ and $V/(V-I)$ CMDs for the stars in the Sh2-311, H18, H19 and nearby reference field regions. The large circles represent the data taken from the literature. Other symbols are the same as in Fig. \ref{op_tcd}. The dashed curve in panel (a) represents the probable background (Norma-Cygnus) population at a distance of $\sim$13.9 kpc, whereas the dashed curve in (b) indicates a foreground population towards the direction of H18  at a distance of $\sim$4.67 kpc. The continuous curves represent the 1 Myr isochrone for Z=0.02 by {\color{red} Marigo et al. (2008)} adjusted for the distance and $E(B-V)$ of the respective clusters (see text). The star shown by an open square in the H19 $(U - B)/(B - V)$ TCD (cf. Fig. \ref{op_tcd}(c)) and panel (c) is probably a foreground star (cf. Section \ref{op_cmd1})}
\end{figure*}

The optical CMDs were used to derive the clusters$^{\prime}$ age, distance, etc. The $V/(B - V)$ and $V/(V - I)$ CMDs of stars within the cluster regions (as discussed in Section \ref{struct_clus}) and the CMD for the reference field (as shown in Fig. \ref{op_cc_image}) are shown in Fig. \ref{op_cmd}. The contamination due to field population is apparent in the CMDs of the cluster regions. The CMDs of the cluster regions also show a significant number of stars towards the right-hand side of the ZAMS. The individual CMDs are discussed below;

{\it Sh2-311:} The brightest star HD 64315 is the ionization source of the region. The upper MS is poorly populated. In Fig. \ref{op_cmd}a, using $E(B - V)_{min}$ = 0.55 mag and $A_V = 3.1$$\times$$E(B - V)$, we visually fitted the theoretical isochrone of log age = 6 (1 Myr) and Z = 0.02 by {\color{red} Marigo et al. (2008)} to the observed sequence. We estimated a distance modulus of $(m - M_V) = 15.2 \pm 0.2$ mag, which corresponds to a distance of 5.0 $\pm$ 0.4 kpc. This is in agreement with the kinematic distance of 5.3 kpc estimated by {\color{red} Balser et al. (2011)} as well as the distance estimated from the spectroscopic observations of star \#A (\cf Section \ref{spec_cla}). There seems to be a background population with  $E(B - V)$ $\sim$0.95 mag at a distance of $\sim$13.9 kpc, which is shown by dashed curve.

{\bf H18:} The CMD of the H18 region (Fig. \ref{op_cmd}b)  shows a complicated distribution of stars.  Adopting $E(B - V)_{min}$ = 0.60 mag the visually fitted isochrones for 1 Myr (Z=0.02) by {\color{red} Marigo et al. (2008)} (continuous line in Fig. \ref{op_cmd}b) to the observations yields $V - M_V$ = 17.1 $\pm$ 0.2 mag, suggesting a distance of 11.2 $\pm$ 1.0 kpc. The cluster's MS is well defined down to  $V$ $\sim$17 mag. The stars fainter than V  $\sim$17 mag  seems to be pre-main-sequence (PMS) stars. The foreground population (dashed line) with $E(B - V)$ = 0.25 mag is found to be located at 4.7 $\pm$ 0.2 kpc. {\color{red} V\'{a}zquez et al. (2010)} have found distances for H18(2) and the foreground population [H18(1)] as 9.5$-$11.4 kpc and 4.5$\pm$0.4 kpc, respectively, which are in fair agreement with the values estimated in the present work.

{\bf H19:}  Fig. \ref{op_cmd}c shows the CMD for the H19 cluster region. The CMD is less complex as compared to the Sh2-311 and H18  regions and show a well defined MS. Using the  $E(B - V)_{min}$ = 0.40 mag, a visual fit to the observations yield $(m - M_V)$ = 15.0 $\pm$ 0.2 mag, which corresponds to a distance of 5.7 $\pm$ 0.4 kpc. {\color{red} V\'{a}zquez et al. (2010)} have pointed out that the brightest stars in the region are not cluster members. The brightest star ($V$=10.931 mag and $(B - V)$ = 0.028 mag) of the region  is located in the outskirts of the region. Its location in $(U - B)/(B - V)$ TCD (\cf Fig. \ref{op_tcd}c, open square) manifests that it should be a field star with $E(B-V)$ $\simeq$ 0.06 mag. The other bright star located in the central region ($V$= 11.050 mag and $(B - V)$ = 0.028 mag) was  described as  the possible exciting source of the region  by {\color{red} Moreno-Corral et al. (2002)}. They have assigned a spectral type of B1.5V on the basis of the $Q$ method (cf. {\color{red} Johnson \& Morgan, 1953}). Its location in $(U - B)/(B - V)$ TCD (\cf Fig. \ref{op_tcd}c, open circle) suggests that it could be a MS star with $E(B - V)$ $\simeq$ 0.40 mag. {\color{red} Moreno-Corral et al. (2002)} have estimated the distance of H19 as 5.2 $\pm$ 0.4 kpc, which is in fair agreement with that estimated in the present study.

\section{\label{ysos_iden} Identification and distribution of YSOs}

Due to field star contamination, the identification of YSOs in clusters/SFRs by using optical data is rather difficult. However,  NIR data are very useful tools to study the nature of young stellar population within the clusters/SFRs. YSOs exhibit strong infrared (IR) excess due to the presence of disks and envelopes, hence  can be identified using the NIR and MIR observations. In general YSOs are distributed in three regions of the NIR-TCD. Objects with accretion disks (Class II) fall on the classical T-Tauri (CTT) locus or along the reddened CTT locus, and objects whose emission is dominated by infalling envelope, known as Class I objects fall redwards of the reddened CTT locus. Stars having inner disk disappeared (Class III) are found in the region corresponding to the reddened main-sequence stars ({\color{red} Meyer et al., 1997}).

\begin{figure*}[!htbp]
\centering
\includegraphics[trim=1cm 4cm 0cm 0cm, clip, height=10cm]{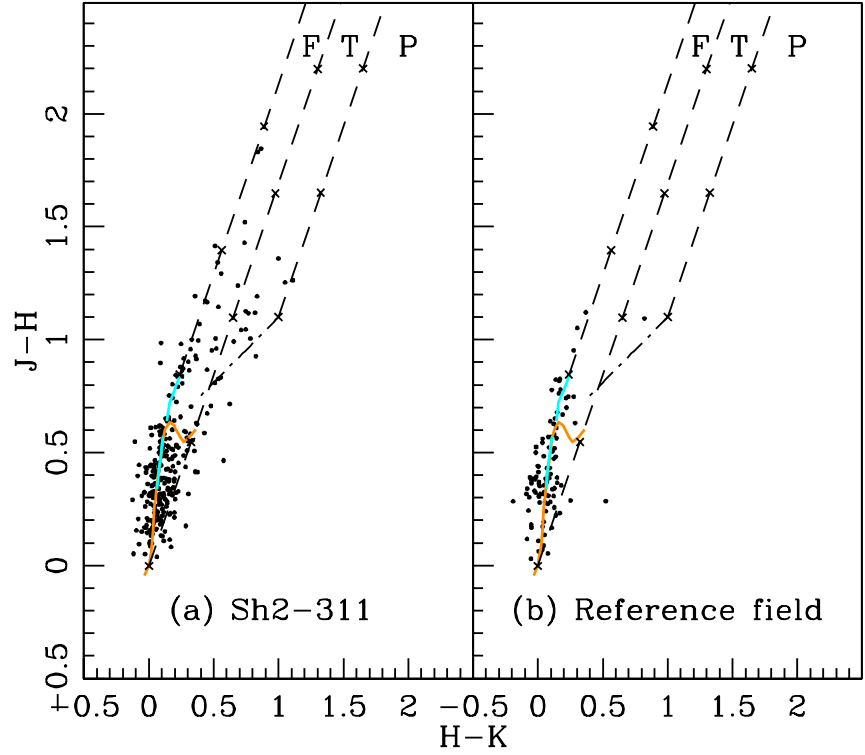}
\caption{\label{nir_tcd} NIR-TCDs for the stars in the Sh2-311 region (left panel) and the field region (right panel). The orange-continuous and cyan-dashed curves represent the unreddened MS and giant branch ({\color{red}Bessell \& Brett, 1988}), respectively. The dot-dashed lines indicate the locus of unreddened CTTSs ({\color{red}Meyer et al., 1997}). The parallel dashed lines are the reddening vectors drawn from the tip (spectral type M4) of the giant branch (left reddening line), from the base (spectral type A0) of the MS branch (middle reddening line), and from the tip of the intrinsic CTTS line (right reddening line) ({\color{red}Cohen et al., 1981}). The crosses on the reddening vectors show an increment of $A_V$ = 5 mag.} 
\end{figure*}

We have used 2MASS NIR photometric data to plot the NIR TCD (cf. Fig. \ref{nir_tcd}) in order to identify the YSOs in the selected regions. The orange-continuous and cyan-dashed curves represent the unreddened MS and giant branches (\cf {\color{red} Bessell \& Brett, 1988}), respectively. The dotted-dashed line indicates the locus of unreddened T-Tauri stars (TTSs) ({\color{red} Meyer et al., 1997}).  The parallel dashed lines are the reddening vectors drawn from the tip (spectral type M4) of the giant branch (left reddening line), from the base (spectral type A0) of the MS branch  (middle reddening line) and from the tip of the intrinsic TTS line (right reddening line). The extinction ratios $A_J /A_V$ = 0.265, $A_H /A_V$ = 0.155 and $A_K /A_V$ = 0.090 have been taken from {\color{red} Cohen et al. (1981)}.  Since the extinction ratios by  {\color{red} Cohen et al (1981)} and the TTS locus by {\color{red} Meyer et al. (1997)}  are in the California Institute of Technology (CIT) system, the $JHK_s$ data were transformed from 2MASS system to the CIT system using the relations given on the 2MASS web site\footnote{http://www.astro.caltech.edu/~jmc/2mass/v3/transformations/}. All the curves and lines are also in the CIT system. The sources located in the `F' region (\cf left panel of Fig. \ref{nir_tcd}) could be either field stars (MS stars, giants) or Class III and Class II sources with small NIR-excesses, whereas the sources distributed in the `T' and `P' regions are considered to be mostly CTTSs or Class II objects with relatively large NIR-excesses and likely Class I objects, respectively (for details see \eg, {\color{red} Pandey et al., 2008}). It is worthwhile to mention that {\color{red} Robitaille et al. (2006)} have shown that there is a significant overlap between protostars and CTTSs. The NIR TCD of the Sh2-311 region  (left panel of Fig. \ref{nir_tcd}) indicates that a few sources are located in the `T' region. A comparison of the NIR-TCD of the Sh2-311 region with that of the NIR-TCD of the nearby reference region (right panel of Fig. \ref{nir_tcd}) suggests that the 8 sources lying in the `T' region could be CTTSs/Class II sources. The NIR and optical data for the identified YSOs are given in Table \ref{ysos_op_data}. 

 Using the {\it Spitzer} MIR observations (3.6 $\mu$m, 4.5 $\mu$m, 5.8 $\mu$m, 8.0 $\mu$m and 24 $\mu$m) for the Sh2-311 region {\color{red} Snider et al. (2009)} have identified 45 YSOs in the region. Twenty-six YSOs have optical counterparts and the optical data for 26 YSOs are given in Table \ref{op_cntp_spit}. The optical data of the YSOs have been used in the ensuing section to estimate the age and mass of the YSOs using the $V/(V-I)$ CMDs.

\subsection{CMDs of YSOs: age and  mass estimation}

\begin{figure*}[!htbp]
\centering
\includegraphics[width=10cm]{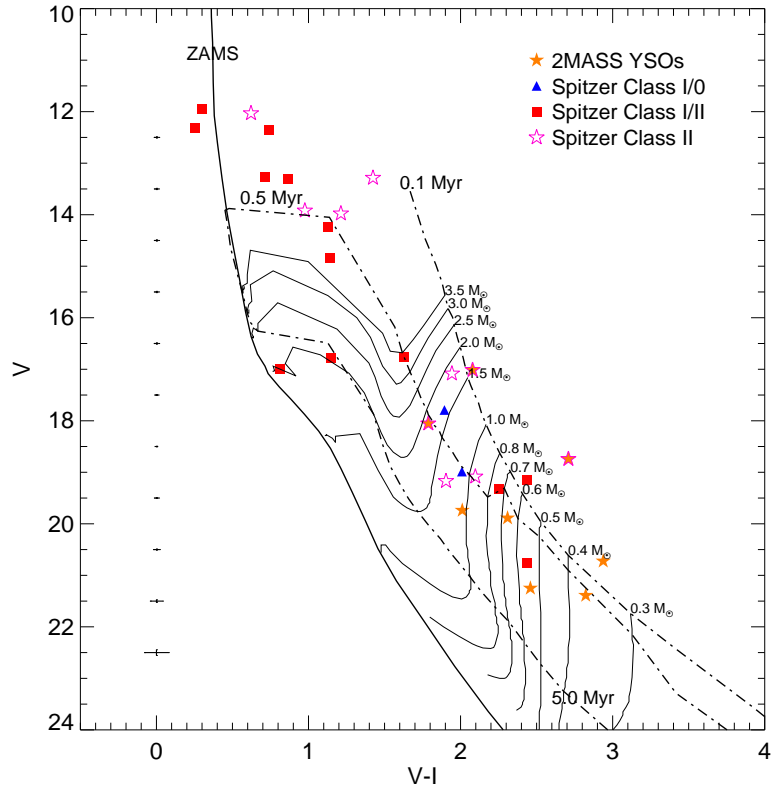}
\caption{\label{op_cmd_ysos} $V/(V-I)$ CMD for the identified YSOs (Class I/0, Class I/II, Class II and NIR-excess sources, shown with blue filled triangle, red filled squares, magenta open stars and orange filled stars, respectively). The PMS isochrones for 0.1, 0.3, 0.5, 1, and 5 Myr by {\color{red}Siess et al. (2000)} and the isochrone for 1 Myr by  {\color{red}Marigo et al (2008)} are drawn as dot-dashed and continuous curves, respectively. The evolutionary tracks by {\color{red} Siess et al. (2000)} for various masses are indicated by thin curves.  All the isochrones and evolutionary tracks are corrected for the Sh2-311 region's distance and average reddening. The typical errors in $V$ and $(V-I)$ are shown on the left side of the figure.}
\end{figure*}

The present deep optical photometry yields  $V$ and $I$ counterparts for 2 Class I/0, 13 Class I/II, 11 Class II and 8 NIR-excess sources. The distribution  of these candidate YSOs  on the $V/(V - I)$ CMD  is shown in Fig. \ref{op_cmd_ysos}. In Fig. \ref{op_cmd_ysos}, 1 Myr isochrone, which is practically identical to the ZAMS (the thick solid curve), by {\color{red} Marigo et al. (2008)} and PMS isochrones by {\color{red} Siess et al. (2000)} for age 0.1, 0.5 and 5 Myr are also shown. The ZAMS and isochrones are shifted for the distance of 5.0 kpc and average reddening \ie $E(B - V)$ = 0.55 mag. The average photometric uncertainty in $V$ magnitudes and $(V - I)$ colors estimated by using the ADDSTAR routine (\cf Section \ref{opt_phot}) are shown in the left-hand side of the figure. The distribution of YSOs on the $V/(V - I)$ CMD can be used  to estimate the approximate ages of YSOs. It is evident from Fig. \ref{op_cmd_ysos} that the majority of the YSOs in the Sh2-311 region are distributed between the PMS isochrones of age 0.1 and 1 Myr. However, the photometric errors,  differential reddening, binary population etc. can cause a spread in the CMD.  Fig. \ref{op_cmd_ysos} also shows the PMS evolutionary tracks (thin solid curves) by {\color{red} Siess et al. (2000)} for various mass bins. The value of mass for each track is given towards its right, which indicates that the PMS sources have masses in the range between 0.4 and 3.5 \msun. 

 The age and mass of each YSO have been estimated using the $V/( V- I)$ CMD, as discussed by {\color{red} Pandey et al.  (2008)} and {\color{red} Chauhan et al.  (2009)} and are given in Table \ref{age_mas}.  Here we would like to point out that the estimation of the ages and masses of the PMS stars by comparing their locations in the CMDs with the theoretical isochrones is prone to random as well as systematic errors ({\color{red} Hillenbrand, 2005; Hillenbrand et al., 2008; Chauhan et al., 2009; Chauhan et al., 2011}). The effect of random errors due to photometric errors and reddening estimation in determination of ages and masses can be estimated by propagating the random errors to their observed estimation by assuming normal error distribution and using the Monte-Carlo simulations (\cf {\color{red} Chauhan et al., 2009}). The systematic errors could be due to the use of different PMS evolutionary models and the error in distance estimation etc. Since we are using evolutionary models by {\color{red} Siess et al.  (2000)} to estimate the age of all the YSOs in the region, we presume that the age estimation is affected only by the random errors. The presence of binaries may also introduce errors in the age determination. Binarity will brighten the star, consequently the CMD will yield a lower age estimate. In the case of an equal mass binary we expect an error of 50$-$60 $\%$ in the PMS age estimation. However, it is difficult to estimate the influence of binaries/variables on mean age estimation as the fraction of binaries/variables is not known. While studying the TTSs in the \hii  region IC 1396, {\color{red} Barentsen et al.  (2011)} pointed out that the number of binaries in their sample of TTSs could be very low as close binaries loose their disk significantly faster than single stars (\cf {\color{red} Bouwman et al., 2006}). Estimated ages and masses of the YSOs range from 0.1 to 8.6 Myr and 0.40$-$6.5 \msun  respectively, which are comparable with the lifetime and masses of TTSs. The age spread indicates a non-coeval star formation in this region. The mean age of the YSOs, (excluding 4 stars having $V$ $<$ 13 and the star lying near the MS with $V$ = 17.0 mag) is found to be $\sim$0.9 Myr.

\begin{figure*}[!htbp]
\centering
\hbox{
\includegraphics[width=7.5cm,angle=270]{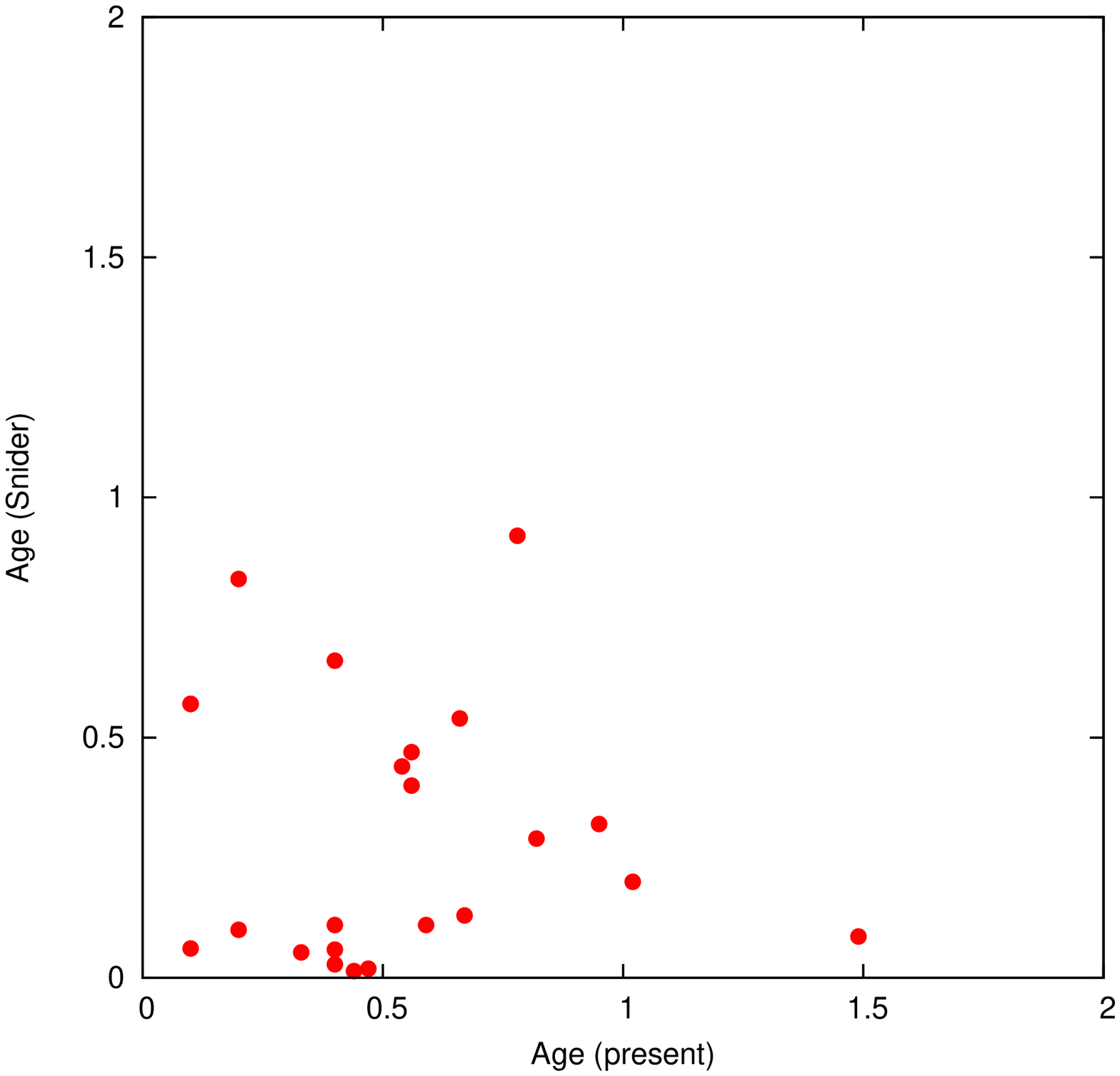}
\includegraphics[width=7.5cm,angle=270]{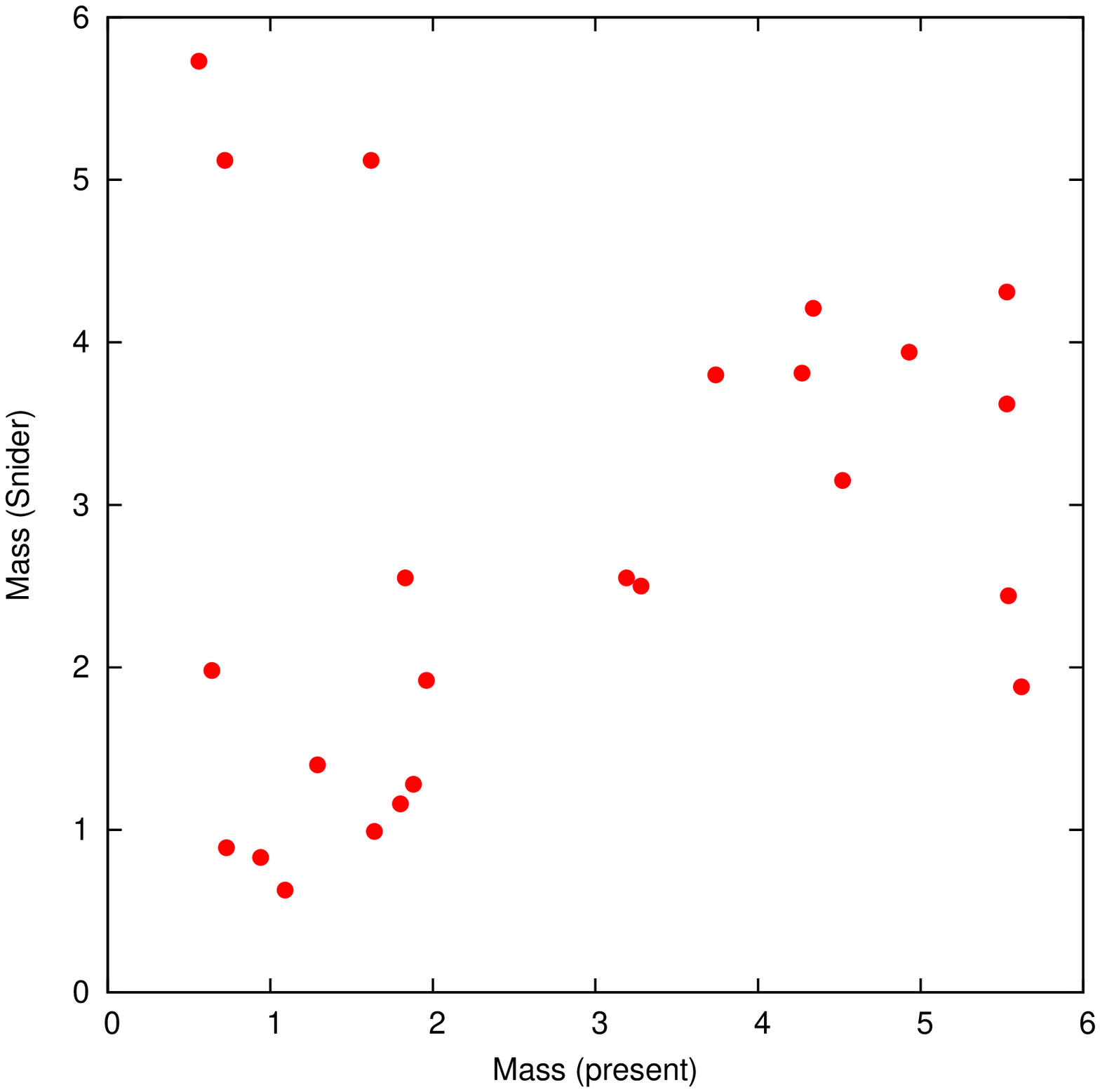}
}
\caption{\label{age_comp} Left panel: Comparison of age obtained in the present study with that reported by {\color{red} Snider et al. (2009)}. Right panel: Comparison of mass obtained in the present study with that reported by {\color{red} Snider et al. (2009)}.}
\end{figure*}

 {\color{red} Snider et al. (2009)} have estimated the age and mass of 45 identified YSOs  using the SEDs generated with the help of grid of models and fitting tools of {\color{red} Robitaille et al. (2006, 2007)}. The age and mass by {\color{red} Snider et al. (2009)} are compared with the present estimation based on CMD analysis in Fig. \ref{op_cmd_ysos}.  There seems no correlation between the two age estimates (\cf Fig. \ref{age_comp}, left panel), however the present age estimation in general yield higher value for age than the {\color{red} Snider et al. (2009)} estimate. This is in contradiction as the CMDs with smaller value of  distance (4.1 kpc) as used by Snider, will make stars relatively older.  Here it is worthwhile to mention that the age estimates based on the $V/(V - I)$ CMDs can be considered more reliable.  Fig. \ref{age_comp}, right panel compares the mass estimates by {\color{red} Snider et al. (2009)} with those obtained in the present study. Though the mass  estimations in general show a scattered distribution, however both the mass estimates seems to be correlated with each other.  However one should expect relatively less mass estimate for a star if we assume relatively less distance for the star.

The ages of young clusters/\hii regions can also be derived from the post-main-sequence age of their most massive members. The most massive member of the Sh2-311 region  is HD 64315, which is an O6-O7V star and seems to be on the MS. This star is considered to be the main ionizing source of the region. Hence, the age of the \hii region Sh2-311 should be younger or of the order of the MS lifetime of the O6-O7V  star. \ie $\sim$5 Myr ({\color{red} Meynet et al., 1994}). Thus, we can put an upper limit to the age of Sh2-311 as $\sim$5 Myr. However, this has to be taken as an approximate estimate as the low-mass members of the region could have been formed prior to HD 64315.

\section{Star formation history of the region}

Massive stars in SFRs significantly affect the entire region. Energetic UV radiation and stellar wind from OB type stars as well as the blast waves from a supernova explosion can evaporate and disperse the surrounding clouds, consequently terminating star formation in the region.  Alternatively, they may play a constructive role in inducing star formation activities in the neighborhood. The expanding ionization/shock fronts caused by O type stars squeeze the molecular clouds and incite a sequence of next generation of star formation ({\color{red} Elmegreen \& Lada, 1977}).  The distribution of YSOs and the morphological details of the environment around the massive star(s) are frequently used to probe the star formation history of the region (\eg, {\color{red} Pandey et al., 2013a; Pandey et al., 2013b} and references therein).

Sh2-311 is a rather spherical \hii region around the ionizing star and is partially surrounded by a dust ring, as revealed by MIR emission in the MSX A-band observations centered at 8.28 $\mu$m (see Fig. \ref{op_cc_image}). The far-ultraviolet (FUV) radiation can escape from the \hii region and penetrate the surface of molecular clouds leading to the formation of a photo-dissociation region (PDR) in the surface layer of the clouds. Polycyclic aromatic hydrocarbons (PAHs) within the PDR are excited by the UV photons, re-emitting their energy at MIR wavelengths, particularly between 6 and 10 $\mu$m.  The MSX A-band includes several discrete PAH emission features (\eg, 6.2, 7.7, and 8.6 $\mu$m) in addition to the contribution from the thermal continuum component from hot dust. The ring of PAH emission surrounds the ionized region and indicates the interface between the ionized and molecular gas (\ie PDR). The absence of 8.28 $\mu$m emission in the interior of the \hii region is interpreted as the destruction of PAH molecules by intense UV radiation of the ionizing star. 

\begin{figure*}[!htbp]
\centering
\resizebox{9cm}{9cm}{\includegraphics{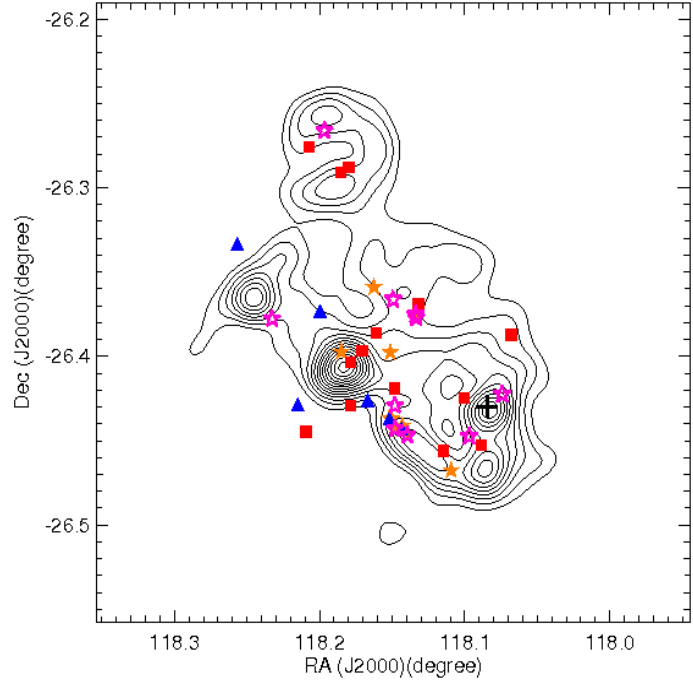}}
\caption{\label{spatial_dist} Spatial distribution of YSO candidates in the Sh2-311 region. Filled star symbols represent the YSOs identified by using the 2MASS NIR data. Filled triangles, filled squares and open star symbols represent the Class I/0, Class I/II and Class II sources, respectively by {\color{red}Snider et al. (2009)} on the basis of MIR data. The 8.28 $\mu$m MSX A-band contours are also overplotted. The ionizing source (HD 64315) is shown by a {\huge +} symbol.}
\end{figure*}

 In the case of triggered star formation we would expect that the spatial distribution of YSOs should be correlated with compressed gas or ionization fronts. Fig. \ref{spatial_dist} over-plots spatial distribution of YSOs along with the MSX A-band MIR contours. The MIR contours partially covers the ionized region around the ionizing source HD 64315 of Sh2-311. It is interesting to note that a significant number of YSOs are distributed along the  MIR contours. Fig. \ref{spatial_dist} suggests that the YSOs could be associated  with the compressed gas/ionization front. As pointed out by {\color{red} Snider et al. (2009)} the radiation driven implosion (RDI) process can be ruled out as we do not see aligned distribution of stars expected in the  case of RDI (\eg, as shown by {\color{red} Chauhan et al., 2009}).  {\color{red} Snider et al. (2009)} favored the shock front traveling in advance of the ionization front as the triggering mechanism over the ``collect and collapse" scenario because they do not see strong evidence for regularly spaced protostars forming around the edge of the \hii region.  However, they mentioned that it was  still plausible that some YSOs in this region formed by the ``collect and collapse"  mechanism as well. In our opinion the spatial distribution of YSOs shown in Fig. \ref{spatial_dist} reveals a systematic distribution of YSOs around the ionization source (HD 64315) of the Sh2-311 region. 

The maximum MS life time of the ionizing source could be $\sim$5 Myr, the spherical morphology of the ionized region Sh2-311, partially surrounded by the dust emission (see Fig. \ref{spatial_dist}), and the association of YSOs (having age $\lesssim$1 Myr) with the dust region suggest triggered star formation in the Sh2-311 region could also be due to the collect and collapse process. 

\section{Conclusions}
This paper presents a deep optical study of the stellar content of the \hii region Sh2-311. Three stellar groups, namely Sh2-311, H18 and H19 are located in this region.  We have made an attempt to distinguish the stellar content of these individual regions as well as to re-determine their fundamental parameters such as distance, reddening, age, onto the basis of a new and more extended optical and infrared photometric data set. The reddening $E(B - V)$ in these regions is found to be variable indicating the presence of the remains of the parental molecular cloud in the region. Sh2-311 and H19 are found to be located in the Perseus arm at a distance of 5.0 $\pm$ 0.4 and 5.7 $\pm$ 0.4 kpc respectively, which suggests that Sh2-311 and H19 could be physically associated as proposed by {\color{red} Moreno-Corral et al. (2002)}. There seems to be a background population at a distance of 13.9 kpc towards the direction of Sh2-311. The H18 cluster is found to be situated at a distance of 11.2 $\pm$ 1.0 kpc with a contaminating foreground population due to the Perseus arm. The ages and masses of a majority of the PMS sources are found to be in the range 0.1 - 1 Myr and 0.4 - 3.5 \msun, respectively, which are comparable with the ages and masses of TTSs and their siblings.

Our spectroscopic observations suggest that the ionizing source of Sh2-311 (HD 64315) is an O6 $\pm$ 1V type star. Sh2-311 is a spherical \hii region around the ionizing source and is surrounded by a dust ring as revealed by MSX A-band 8.28 $\mu$m observations. A significant number of YSOs are distributed along the 8.28 $\mu$m emission. 

The age of the ionizing source ($\sim$5 Myr) as well as the morphology of Sh2-311, the spatial distribution of YSOs and the average age $\sim$1 Myr of YSOs, indicate for a triggered star formation in the outer region of Sh2-311. A deep NIR observations of the region may be useful to study the young stellar content of the region. 

\section*{Acknowledgments}
The authors are grateful for useful comments from an anonymous referee that improved the article. The authors are thankful to the staff of HCT (operated by Indian Institute of Astrophysics, Bangalore) for their assistance and support during the observations. The WFI observations used in the present work have been taken from the ESO data archive. R.K.Yadav is highly thankful to Prof. Frank Valdes for his help and fruitful discussion during WFI data reduction. This research  has made use of the Simbad and Vizier databases maintained at CDS, Strasbourg, France; NASA's  Astrophysics Data System Bibliographic Services; WEBDA database operated at the Department of Theoretical Physics and Astrophysics of the Masaryk University; and NASA/IPAC Infrared Science Archive operated by the Jet Propulsion Laboratory, California Institute of Technology. We also acknowledge STARLINK's STILTS package for command-line processing of tabular data. The STARLINK project was a long running UK project supporting astronomical data processing.

\clearpage


\begin{table*}
\caption{Log of observations.}
\label{obslog}
\begin{tabular}{ccccc}
\hline
\hline
\ra & \dec & Date of &Filter &No. of frames$\times$ Exp. time (sec)\\
        &          &    observation  &          &  \\
\hline
{\it \bf WFI$^1$}&     &  &   &                        \\
&   & 2002-10-29 & $ I $      &(7x20)                  \\
07:52:38.88& -26:22:20.10  & 2002-10-30 & $ UBV $    &(7x60,7x30,7x30)        \\
&   & 2002-12-07 & $ VI  $    &(2x1200, 2x900)         \\

\hline
{\it \bf HCT$^2$} &                   &  &        \\
07:52:20.28 &-26:25:46.71& 2012-12-14 &  & 1x900  \\
07:52:16.84 &-26:23:32.73& 2012-12-14 &  & 1x1500 \\
07:52:13.68 &-26:24:00.20& 2012-12-14 &  & 1x1500 \\
\hline

\end{tabular}

$^1$  Wide Field Imager on MPG/ESO 2.2m Telescope.\\
$^2$  Himalayan Chandra Telescope (HCT), Hanle.\\
\end{table*}

\begin{table*}
\caption{\label{col_coef} The color coefficients and zero-point coefficients.}
\begin{tabular}{ccc}
\hline
\hline
Parameters  & Short exposure  & Long exposure  \\
\hline
{\bf Color Coefficients}   &&  \\
$m_1$  &    0.982$\pm$0.022         &      .....                   \\
$m_2$  &    1.418$\pm$0.034         &      .....                  \\
$m_3$  &    0.900$\pm$0.011         &   0.906$\pm$0.008             \\
$m_4$  &   -0.067$\pm$0.006         &  -0.064$\pm$0.004             \\

{\bf Zero-points Coefficients}  && \\
$c_1$  &    -3.060$\pm$0.072        &    .....                    \\
$c_2$  &     0.711$\pm$0.008        &    .....                    \\
$c_3$  &     0.571$\pm$0.004        &   0.893$\pm$0.002             \\
$c_4$  &    -1.345$\pm$0.005        &  -1.536$\pm$0.005             \\
\hline
\end{tabular}
\end{table*}

\begin{table*}
\scriptsize{
\caption{\label{op_data} $UBVI_c$ photometric data of stars in Sh2-311 region. The complete table is available in electronic form only.}
\begin{tabular}{ccccccc}
\hline
\hline
ID & \ra & \dec & $V\pm\sigma$ & $(U-B)\pm\sigma$ & $(B-V)\pm\sigma$ & $(V-I)\pm\sigma$\\
\hline
 61   &07:53:49.01&-26:14:09.67&12.233 $\pm $0.020&0.397 $\pm$ 0.002& 0.083 $\pm$ 0.002&0.638 $\pm$ 0.005\\  
 62   &07:52:25.84&-26:32:19.51&12.242 $\pm $0.008&1.131 $\pm$ 0.004& 0.941 $\pm$ 0.005&1.209 $\pm$ 0.005\\  
 63   &07:52:50.46&-26:22:39.32&12.255 $\pm $0.018&0.396 $\pm$ 0.012& 0.293 $\pm$ 0.009&0.519 $\pm$ 0.010\\  
 64   &07:53:49.65&-26:22:41.31&12.255 $\pm $0.014&0.495 $\pm$ 0.003& 0.103 $\pm$ 0.003&1.096 $\pm$ 0.009\\  
 65   &07:53:17.36&-26:31:40.17&12.266 $\pm $0.008&0.006 $\pm$ 0.019&-0.293 $\pm$ 0.009&0.229 $\pm$ 0.028\\  
\hline
\end{tabular}
}
\end{table*}

\begin{table*}
\scriptsize{
\caption{Comparison of present photometry with that available in the literature. The $\Delta$ represents the difference between the present and literature value.}
\label{photdata}
\begin{tabular}{|l|cccc} 
\hline
\backslashbox{Photometry from literature}{$\Delta$ Magnitude/$\Delta$ Color} & $\Delta V \pm\sigma$ & $\Delta (B-V)\pm\sigma$ & $\Delta (U-B) \pm\sigma$ & $\Delta (V-I) \pm\sigma$ \\
\hline
{\bf Photoelectric photometry} &&&&\\

Lod{\'e}n (1966)                   & -0.016$\pm$ 0.081& -0.041$\pm$ 0.071& 0.030$\pm$ 0.108&                 \\
Moffat \& FitzGerald (1974)    &  0.099$\pm$ 0.170&  0.086$\pm$ 0.119& 0.042$\pm$ 0.141&                 \\
Feinstein \& V\'{a}zquez (1989)        &  0.011$\pm$ 0.059& -0.007$\pm$ 0.060& 0.000$\pm$ 0.054& 0.019$\pm$ 0.039\\
                                                                                                          \\     
{\bf CCD photometry} &&&&\\                                                                              
Labhardt et al. (1992)         & -0.036$\pm$ 0.052& -0.005$\pm$ 0.072&                 &                 \\
Munari \& Carraro (1996)       & -0.027$\pm$ 0.054& -0.080$\pm$ 0.033& 0.248$\pm$ 0.150& 0.020$\pm$ 0.023\\
Munari et al. (1998)           &  0.040$\pm$ 0.084& -0.007$\pm$ 0.102& 0.046$\pm$ 0.105& 0.069$\pm$ 0.117\\
Moreno-Corral et al. (2002)    &  0.048$\pm$ 0.415&  0.023$\pm$ 0.442& 0.175$\pm$ 0.361& 0.058$\pm$ 0.144\\
Vazquez et al. (2010)          &  0.001$\pm$ 0.078&  0.020$\pm$ 0.105& 0.055$\pm$ 0.161& 0.004$\pm$ 0.090\\
\hline
\end{tabular}
}
\end{table*}

\begin{table*}
\caption{Completeness Factor}
\label{cft}
\begin{tabular}{ccccc}
\hline
\hline
V range & Sh2-311 & Haffner 18 & Haffner 19  & Field \\
(mag)  &  r $\le$ 3.${\arcm}$0 & r $\le$ 1.${\arcm}$7  & r $\le$ 2.${\arcm}$0 & r $\le$ 3.${\arcm}$0\\
\hline
12-13 &  1.00  &  1.00    &  1.00 &  1.00 \\
13-14 &  1.00  &  1.00    &  1.00 &  1.00 \\
14-15 &  1.00  &  1.00    &  1.00 &  1.00 \\
15-16 &  1.00  &  1.00    &  1.00 &  1.00 \\
16-17 &  0.98  &  0.98    &  0.98 &  0.96 \\
17-18 &  0.98  &  0.98    &  0.97 &  0.98 \\
18-19 &  0.96  &  0.97    &  0.97 &  0.96 \\
19-20 &  0.95  &  0.95    &  0.95 &  0.95 \\
20-21 &  0.89  &  0.90    &  0.95 &  0.89 \\
21-22 &  0.87  &  0.85    &  0.88 &  0.87 \\
22-23 &  0.59  &  0.58    &  0.59 &  0.59 \\
\hline
\end{tabular}
\end{table*}

\begin{table*}
\scriptsize 
\caption{\label{spec_mem} $UBVIJHK$ photometric data of 3 stars in Sh2-311 region studied spectroscopically. The photometric data have been taken from Simbad and Webda.}
\begin{tabular}{cccccccccccc}
\hline
\hline
Star & \ra & \dec & $U$ & $B$ & $V$ & $R$ & $I$ & $J$ & $H$ & $K$ & Sp. type\\
\hline
 A   & 07:52:20.28  & -26:25:46.70 &  8.67 &  9.37  &   9.19 &  9.07  & 8.86  & 8.61& 8.55 & 8.49 &  O6V\\
 B   & 07:52:16.84  & -26:23:32.68 & 11.51 & 11.51  &  11.32 & 11.64  &11.08  &10.95&10.90 &10.85 &  B6V\\
 C   & 07:52:13.66  & -26:24:01.23 & 11.68 & 11.68  &  11.60 &  ...   & ...   &11.35&11.35 &11.26 &  B7V\\ 
\hline
\end{tabular}
\end{table*}


\hspace{-2cm}
\begin{table*}
\tiny
\caption{\label{ysos_op_data} Optical counterparts of YSOs identified on the basis of NIR-TCD.}
\begin{tabular}{cccccccc}
\hline
S. No. & \ra & \dec & $V\pm\sigma$ &  $(V-I)\pm\sigma$ & $(J-H)\pm\sigma$ & $(H-K)\pm\sigma$ &  $K\pm\sigma$ \\
\hline
1 & 07:52:33.47 & -26:26:48.04 & 17.014$\pm$0.005 & 2.079$\pm$0.004 & 1.253$\pm$0.029 & 1.050$\pm$0.034 & 10.550$\pm$0.026 \\
2 & 07:52:32.15 & -26:22:29.83 & 18.053$\pm$0.006 & 1.788$\pm$0.005 & 1.042$\pm$0.045 & 0.712$\pm$0.024 & 13.261$\pm$0.041 \\
3 & 07:52:44.41 & -26:23:50.59 & 19.738$\pm$0.006 & 2.011$\pm$0.007 & 1.192$\pm$0.073 & 0.833$\pm$0.047 & 13.798$\pm$0.053 \\
4 & 07:52:34.32 & -26:26:27.73 & 19.888$\pm$0.013 & 2.309$\pm$0.007 & 1.005$\pm$0.106 & 0.783$\pm$0.097 & 13.809$\pm$0.064 \\
5 & 07:52:39.04 & -26:21:32.58 & 20.724$\pm$0.027 & 2.938$\pm$0.023 & 1.045$\pm$0.065 & 0.746$\pm$0.047 & 13.884$\pm$0.051 \\
6 & 07:52:36.32 & -26:23:51.06 & 21.250$\pm$0.020 & 2.459$\pm$0.016 & 1.119$\pm$0.085 & 0.819$\pm$0.069 & 14.101$\pm$0.064 \\
7 & 07:52:36.08 & -26:26:13.43 & 21.389$\pm$0.025 & 2.823$\pm$0.025 & 1.126$\pm$0.081 & 0.747$\pm$0.073 & 13.876$\pm$0.055 \\
8 & 07:52:26.21 & -26:28:02.45 & 21.475$\pm$0.058 & 2.847$\pm$0.011 & 0.992$\pm$0.087 & 0.653$\pm$0.099 & 14.165$\pm$0.078 \\
\hline
\end{tabular}
\end{table*}

\begin{table*}
\tiny
\caption{\label{op_cntp_spit} Optical counterparts of YSOs identified by {\color{red} Snider et al. (2009)}.}
\begin{tabular}{cccccccc}
\hline
ID & \ra & \dec &$V$$\pm\sigma$  & $(B-V)$$\pm\sigma$ & $(U-B)$$\pm\sigma$ &  $(V-I)$$\pm\sigma$&  Class\\
({\color{red}Snider et al., }  &           &            &              &                &                &                &                    \\
 {\color{red} 2009})  &           &            &              &                &                &                &                   \\
\hline

      5&07:52:36.43&-26:26:11.33&22.120 $\pm$ 0.047&       .....       &       .....       & 3.100 $\pm$ 0.017 & I/0\\  
      6&07:52:39.98&-26:25:33.22&17.791 $\pm$ 0.004&       .....       &       .....       & 1.894 $\pm$ 0.015 & I/0\\  
     11&07:52:47.88&-26:22:23.02&18.986 $\pm$ 0.018&       .....       &       .....       & 2.009 $\pm$ 0.022 & I/0\\  
     17&07:52:21.28&-26:27:09.74&13.261 $\pm$ 0.010& 0.619 $\pm$  0.003& 0.126 $\pm$  0.004& 0.714 $\pm$ 0.005 &I/II\\  
     18&07:52:24.01&-26:25:28.40&12.358 $\pm$ 0.009& 0.619 $\pm$  0.003& 0.053 $\pm$  0.003& 0.742 $\pm$ 0.005 &I/II\\  
     19&07:52:27.51&-26:27:20.97&17.000 $\pm$ 0.011& 0.642 $\pm$  0.006& 0.465 $\pm$  0.020& 0.813 $\pm$ 0.013 &I/II\\  
     20&07:52:31.68&-26:22:07.24&16.771 $\pm$ 0.041& 0.809 $\pm$  0.019& 0.439 $\pm$  0.026& 1.627 $\pm$ 0.030 &I/II\\  
     21&07:52:35.54&-26:25:07.97&19.158 $\pm$ 0.006&       .....       &       .....       & 2.434 $\pm$ 0.006 &I/II\\  
     22&07:52:36.39&-26:25:56.47&20.759 $\pm$ 0.015&       .....       &       .....       & 2.435 $\pm$ 0.025 &I/II\\  
     23&07:52:38.57&-26:23:10.10&14.246 $\pm$ 0.013& 0.896 $\pm$  0.011&-0.152 $\pm$  0.011& 1.128 $\pm$ 0.010 &I/II\\  
     25&07:52:40.96&-26:23:48.38&14.841 $\pm$ 0.011& 0.791 $\pm$  0.011&-0.064 $\pm$  0.010& 1.144 $\pm$ 0.014 &I/II\\  
     26&07:52:44.60&-26:17:27.22&12.313 $\pm$ 0.007& 0.212 $\pm$  0.005&-0.818 $\pm$  0.001& 0.255 $\pm$ 0.004 &I/II\\  
     27&07:52:42.24&-26:22:51.01&19.324 $\pm$ 0.006&       .....       &       .....       & 2.251 $\pm$ 0.007 &I/II\\  
     28&07:52:42.89&-26:24:12.83&11.940 $\pm$ 0.009& 0.141 $\pm$  0.009&-0.632 $\pm$  0.018& 0.303 $\pm$ 0.010 &I/II\\  
     29&07:52:42.79&-26:25:46.21&16.785 $\pm$ 0.008& 0.840 $\pm$  0.005& 0.690 $\pm$  0.027& 1.146 $\pm$ 0.004 &I/II\\  
     32&07:52:50.24&-26:26:40.96&13.307 $\pm$ 0.008& 0.601 $\pm$  0.002&-0.301 $\pm$  0.003& 0.868 $\pm$ 0.004 &I/II\\  
     33&07:52:17.79&-26:25:20.22&13.977 $\pm$ 0.008& 1.124 $\pm$  0.003& 0.791 $\pm$  0.005& 1.213 $\pm$ 0.005 &  II\\  
     34&07:52:23.24&-26:26:49.90&13.284 $\pm$ 0.010& 1.186 $\pm$  0.003& 1.246 $\pm$  0.004& 1.424 $\pm$ 0.006 &  II\\  
     35&07:52:32.03&-26:22:37.00&19.084 $\pm$ 0.006&       .....       &       .....       & 2.097 $\pm$ 0.006 &  II\\  
     36&07:52:32.15&-26:22:29.83&18.053 $\pm$ 0.006&       .....       &       .....       & 1.788 $\pm$ 0.005 &  II\\  
     38&07:52:33.47&-26:26:48.04&17.014 $\pm$ 0.005&       .....       &       .....       & 2.079 $\pm$ 0.004 &  II\\  
     39&07:52:35.53&-26:25:44.46&13.920 $\pm$ 0.024& 0.356 $\pm$  0.005&-0.528 $\pm$  0.004& 0.976 $\pm$ 0.010 &  II\\  
     40&07:52:35.52&-26:26:34.38&18.742 $\pm$ 0.029&       .....       &       .....       & 2.709 $\pm$ 0.018 &  II\\  
     41&07:52:36.21&-26:25:47.64&19.170 $\pm$ 0.007&       .....       &       .....       & 1.904 $\pm$ 0.016 &  II\\  
     42&07:52:39.14&-26:23:22.44&12.031 $\pm$ 0.004& 0.349 $\pm$  0.004&-0.578 $\pm$  0.011& 0.622 $\pm$ 0.004 &  II\\  
     45&07:52:55.89&-26:22:39.93&17.084 $\pm$ 0.004&       .....       &       .....       & 1.943 $\pm$ 0.005 &  II\\  

\hline
\end{tabular}
\end{table*}

\begin{table*}
\centering
\caption{\label{age_mas} Ages and masses of the YSOs.}
\begin{tabular}{ccccc}
\hline
ID & $V \pm \sigma$ & $(V-I)\pm\sigma$ & Age$\pm\sigma$ & Mass$\pm\sigma$ \\
   &                &                  &  (Myr)         &      \msun      \\
\hline
    1 &  13.261 $\pm$ 0.010 &  0.714 $\pm$ 0.005 &  0.56 $\pm$ 0.19 &  4.93 $\pm$ 1.04\\
    2 &  13.284 $\pm$ 0.010 &  1.424 $\pm$ 0.006 &  0.20 $\pm$ 0.00 &  6.52 $\pm$ 0.04\\
    3 &  13.307 $\pm$ 0.008 &  0.868 $\pm$ 0.004 &  0.66 $\pm$ 0.07 &  4.34 $\pm$ 0.46\\
    4 &  13.920 $\pm$ 0.024 &  0.976 $\pm$ 0.010 &  0.78 $\pm$ 0.34 &  4.27 $\pm$ 0.88\\
    5 &  13.977 $\pm$ 0.008 &  1.213 $\pm$ 0.005 &  0.67 $\pm$ 0.05 &  4.52 $\pm$ 0.26\\
    6 &  14.246 $\pm$ 0.013 &  1.128 $\pm$ 0.010 &  0.95 $\pm$ 0.23 &  3.74 $\pm$ 0.46\\
    7 &  14.841 $\pm$ 0.011 &  1.144 $\pm$ 0.014 &  1.49 $\pm$ 0.26 &  3.19 $\pm$ 0.44\\
    8 &  16.771 $\pm$ 0.041 &  1.627 $\pm$ 0.030 &  0.54 $\pm$ 0.11 &  3.28 $\pm$ 0.08\\
    9 &  16.785 $\pm$ 0.008 &  1.146 $\pm$ 0.004 &  6.46 $\pm$ 0.56 &  1.96 $\pm$ 0.05\\
   10 &  17.014 $\pm$ 0.005 &  2.079 $\pm$ 0.004 &  0.10 $\pm$ 0.00 &  1.62 $\pm$ 0.05\\
   11 &  17.084 $\pm$ 0.004 &  1.943 $\pm$ 0.005 &  0.20 $\pm$ 0.02 &  1.83 $\pm$ 0.09\\
   12 &  17.791 $\pm$ 0.004 &  1.894 $\pm$ 0.015 &  0.33 $\pm$ 0.05 &  1.64 $\pm$ 0.13\\
   13 &  18.053 $\pm$ 0.006 &  1.788 $\pm$ 0.005 &  0.59 $\pm$ 0.08 &  1.88 $\pm$ 0.12\\
   14 &  18.742 $\pm$ 0.029 &  2.709 $\pm$ 0.018 &  0.10 $\pm$ 0.00 &  0.72 $\pm$ 0.01\\
   15 &  18.742 $\pm$ 0.029 &  2.709 $\pm$ 0.018 &  0.10 $\pm$ 0.00 &  0.73 $\pm$ 0.01\\
   16 &  18.986 $\pm$ 0.018 &  2.009 $\pm$ 0.022 &  0.56 $\pm$ 0.07 &  1.09 $\pm$ 0.06\\
   17 &  19.084 $\pm$ 0.006 &  2.097 $\pm$ 0.006 &  0.47 $\pm$ 0.05 &  0.94 $\pm$ 0.06\\
   18 &  19.158 $\pm$ 0.006 &  2.434 $\pm$ 0.006 &  0.10 $\pm$ 0.00 &  0.64 $\pm$ 0.01\\
   19 &  19.170 $\pm$ 0.007 &  1.904 $\pm$ 0.016 &  1.02 $\pm$ 0.16 &  1.29 $\pm$ 0.08\\
   20 &  19.324 $\pm$ 0.006 &  2.251 $\pm$ 0.007 &  0.44 $\pm$ 0.05 &  0.73 $\pm$ 0.05\\
   21 &  19.738 $\pm$ 0.006 &  2.011 $\pm$ 0.007 &  1.16 $\pm$ 0.17 &  1.05 $\pm$ 0.05\\
   22 &  19.888 $\pm$ 0.013 &  2.309 $\pm$ 0.007 &  0.62 $\pm$ 0.05 &  0.66 $\pm$ 0.03\\
   23 &  20.724 $\pm$ 0.027 &  2.938 $\pm$ 0.023 &  0.10 $\pm$ 0.00 &  0.38 $\pm$ 0.01\\
   24 &  20.759 $\pm$ 0.015 &  2.435 $\pm$ 0.025 &  0.82 $\pm$ 0.08 &  0.56 $\pm$ 0.03\\
   25 &  21.250 $\pm$ 0.020 &  2.459 $\pm$ 0.016 &  1.17 $\pm$ 0.14 &  0.54 $\pm$ 0.02\\
   26 &  21.389 $\pm$ 0.025 &  2.823 $\pm$ 0.025 &  0.67 $\pm$ 0.16 &  0.36 $\pm$ 0.01\\
\hline
\end{tabular}
\end{table*}

\end{document}